%% file: models.tex
\documentclass[11pt,a4paper]{article}
\pdfoutput=1

\usepackage{amsmath}
\usepackage{amssymb}
\usepackage{graphicx}
\usepackage{cite}
\usepackage{hyperref}
\usepackage{xcolor}
\usepackage{mathrsfs}
\usepackage{booktabs}
\usepackage{multirow}
\usepackage[utf8]{inputenc}
\usepackage{rotating}
\usepackage{feynmp}
\usepackage{scalerel}

\newsavebox\feynbox
\newlength\tmplength
\newcommand\postfeynlabel[4]{%
  \setlength{\tmplength}{#2\ht#3}%
  \raisebox{#1\tmplength}{%
  \scaleto[1.7ex]{\raisebox{2.33pt}{\}}}{\tmplength}%
  \raisebox{\dimexpr.5\tmplength-.35\ht\strutbox}{#4}}%
}
\begin{document}

\vspace{-2.0cm}
\begin{flushright}
Edinburgh 2016/04\\
OUTP-16-01P \\
\end{flushright}
\vspace{.1cm}

\begin{center}
{\Large \bf The asymptotic behaviour of parton distributions\\[0.2cm]
at small and large $x$}
\vspace{.7cm}

Richard~D.~Ball$^{1}$, Emanuele R.~Nocera$^{2}$ and Juan~Rojo$^{2}$

\vspace{.3cm}{\it 
~$^1$ The Higgs Centre for Theoretical Physics, University of Edinburgh,\\
JCMB, KB, Mayfield Rd, Edinburgh EH9 3JZ, Scotland\\
~$^2$ Rudolf Peierls Centre for Theoretical Physics, 1 Keble Road,\\ 
University of Oxford, OX1 3NP Oxford, UK}
\end{center}   

\vspace{0.1cm}
\begin{center}
  {\bf \large Abstract}
\end{center}

It has been argued from the earliest days of quantum chromodynamics (QCD) 
that at asymptotically small values of $x$ the parton distribution functions 
(PDFs) of the proton behave as $x^\alpha$, where the values of $\alpha$ can be deduced from 
Regge theory, while at asymptotically large values of $x$ the PDFs behave as 
$(1-x)^\beta$, where the values of $\beta$ can be deduced from the Brodsky-Farrar 
quark counting rules. We critically examine these claims by extracting the 
exponents $\alpha$ and $\beta$ from various global fits of parton 
distributions, analysing their scale dependence, and comparing their values 
to the naive expectations. We find that for valence distributions both 
Regge theory and counting rules are confirmed, at least within 
uncertainties, while for sea quarks and gluons the results are less conclusive. 
We also compare results from various PDF fits for the structure function 
ratio $F_2^n/F_2^p$ at large $x$, and caution against unrealistic uncertainty 
estimates due to overconstrained parametrisations.

\clearpage


\input{sec-introduction}

\input{sec-exponents}

\input{sec-comparison}

\input{sec-conclusion}

\section*{Acknowledgements}

J.R. is supported by an STFC Rutherford Fellowship ST/K005227/1 and by an 
European Research Council Starting Grant ``PDF4BSM". The work of J.R. and 
E.R.N. is supported by a STFC Rutherford Grant ST/M003787/1.


\input{sec-appendix}

\bibliography{models}

\end{document}

%% file: sec-introduction.tex
\section{Introduction}
\label{sec:introduction}

An accurate determination of Parton Distribution Functions (PDFs) is an 
essential building block for the precision physics program at the Large Hadron 
Collider (LHC)~\cite{Forte:2013wc,Rojo:2015acz,Ball:2015oha,Butterworth:2015oua,Jimenez-Delgado:2014twa}.
Given current 
limitations in the understanding of nonperturbative Quantum Chromodynamics
(QCD), such a determination is not achievable from first principles.
Instead, PDFs are determined in a global fit to hard-scattering experimental
data~\cite{Ball:2014uwa,Harland-Lang:2014zoa,Dulat:2015mca,Alekhin:2013nda,Abramowicz:2015mha,Accardi:2016qay}, using perturbative QCD to combine 
information from different processes and
scales. In such an analysis, the best-fit values of 
the input PDF parametrisation are obtained by comparing the PDF-dependent 
prediction of a suitable set of physical observables with their measured 
values, and then by minimising a figure of merit which quantifies 
the agreement between the two.

The parametrisation of the PDFs, $xf_i(x,Q_0^2)$, is set at an initial scale 
$Q_0^2$, and is then evolved to any other scale $Q^2$ via DGLAP 
equations~\cite{Gribov:1972ri,Altarelli:1977zs,Dokshitzer:1977sg}. 
The PDF parametrisation should be as general as possible, and in particular
sufficiently smooth and flexible enough to accommodate all of the experimental 
data included in the fit without artificial bias. The kinematic constraint 
that  $xf_i(x,Q_0^2)$ vanishes in the elastic limit $x\to 1$ should also be 
implicit in the parametrisation. Usually, the following {\it ansatz} is adopted
\begin{equation}
xf_i(x,Q_0^2) = A_{f_i}\, x^{a_{f_i}}\, (1-x)^{b_{f_i}}\, \mathscr{F} (x,\{c_{f_i}\})
\,\mbox{,}
\label{eq:parametrisation}
\end{equation}
where $x$ is the parton momentum fraction and $i$ denotes a given quark flavour 
(or flavour combination) or the gluon, and $\mathscr{F} (x,\{c_{f_i}\})$ is a 
smooth function which remains finite both when  $x\to 0$ and $x\to 1$.
The normalisation fractions $A_{f_i}$, the exponents $a_{f_i}$ and 
$b_{f_i}$, and the set of parameters $\{ c_{f_i}\}$ are then determined from the 
data. Some of the $A_{f_i}$ can be fixed in terms of the other fit parameters
by means of the momentum and valence sum rules.
 
The original motivation for Eq.~(\ref{eq:parametrisation}) was the theoretical 
expectation, based on nonperturbative QCD considerations, of a power-law 
behaviour of the PDFs at sufficiently small and large values of $x$.
Specifically, Regge theory~\cite{Regge:1959mz} predicts
\begin{equation}
xf_i(x,Q^2)\xrightarrow{x\to 0} x^{a_{f_i}}
\,\mbox{;}
\label{eq:Regge}
\end{equation}
while the Brodsky-Farrar quark counting rules~\cite{Brodsky:1973kr} predict
\begin{equation}
xf_i(x,Q^2)\xrightarrow{x\to 1} (1-x)^{b_{f_i}}
\,\mbox{;}
\label{eq:countingrules}
\end{equation}
see also Ref.~\cite{Roberts:1990ww,Devenish:2004pb}, and references therein.
Both Regge theory and the counting rules
provide numerical predictions for the values of  the exponents
$a_{f_i}$ and $b_{f_i}$.
In Eq.~(\ref{eq:parametrisation}), the small- and large-$x$ power-law 
behaviours are matched at intermediate $x$ values through the 
function $\mathscr{F}(x,\{c_{f_i}\})$. A number of different 
parametrisations have been used for this function so far, 
ranging from simple polynomials
to more sophisticated
Chebyshev~\cite{Glazov:2010bw,Harland-Lang:2014zoa} and
Bernstein~\cite{Dulat:2015mca}
polynomials and multi-layer neural
networks~\cite{DelDebbio:2004xtd,DelDebbio:2007ee}.

It should be emphasised that Eqs.~(\ref{eq:Regge})-(\ref{eq:countingrules})  
{\it cannot} be derived using perturbative QCD, but rather require other more 
general considerations. For instance, counting rules can be derived 
from Bloom-Gilman duality~\cite{Lepage:1980fj} or using AdS/QCD methods 
in nonperturbative QCD~\cite{Polchinski:2001tt}.\footnote {It has been proved 
that counting rules are rigorous predictions of QCD, modulo calculable 
logarithmic corrections from the behaviour of the hadronic wave function at 
short distances, in the case of large momentum transfer exclusive 
processes~\cite{Lepage:1980fj,Matveev:1972gb}.}
The use of Eqs.~(\ref{eq:Regge})-(\ref{eq:countingrules}) in the input PDF 
parametrisation, Eq.~(\ref{eq:parametrisation}), could therefore lead to 
theoretical bias.
For instance, as we will discuss below, perturbative QCD calculations
predict a logarithmic, rather than a power-like, growth of the PDFs at 
small $x$. Even if Eqs.~(\ref{eq:Regge})-(\ref{eq:countingrules})
were a solid prediction from QCD (which they are not), they
would not be  particularly useful in the context of a global
PDF analysis. First, it is unclear how small or large
$x$ should be in order for the power 
laws~(\ref{eq:Regge})-(\ref{eq:countingrules}) to provide
a reliably enough approximation of the underlying PDFs.
Second, it is unclear at which values of $Q^2$ Regge 
theory and Brodsky-Farrar quark counting rules should apply exactly.
This is a serious limitation, given the non-negligible PDF scale dependence 
around the input parametrisation scale $Q^2\simeq Q_0^2$. 
In principle, the optimal values of $Q^2$ 
should be chosen at the interface between perturbative and nonperturbative 
hadron dynamics, $Q^2\simeq Q_0^2 = Q^2_{\rm in}$. 
It has been shown~\cite{Deur:2016cxb} that $Q_{\rm in}^2\simeq 0.75$ GeV$^2$ 
by matching the high- and low-$Q^2$ behaviour of the
strong coupling $\alpha_s(Q^2)$ as predicted respectively by its 
renormalization group equation in the $\overline{\rm MS}$ scheme and its 
analytic form in the light-front holographic approach.

The aim of this study is to present a methodology to quantify the effective
asymptotic behaviour of PDFs at small and large values of $x$, and then
apply it to compare recent global fits with various
perturbative and nonperturbative QCD predictions.
The paper is organised as follows. In Sec.~\ref{sec:exponents} we 
introduce a definition of the effective PDF exponents, and we use them 
to quantify for which ranges of $x$ and $Q^2$,
if any, PDFs exhibit a power-law behaviour of the form
Eqs.~(\ref{eq:Regge})-(\ref{eq:countingrules}).
Once the asymptotic range has been determined, in Sec.~\ref{sec:comparisons}
we investigate to which extent these exponents, as obtained from global 
PDF fits, are in agreement with the theoretical predictions of their values.
In addition to Brodsky-Farrar quark counting rules, we will also compare the 
global fit predictions with other nonperturbative models of nucleon structure
at large $x$. In principle, this comparison will allow us to discriminate 
among models, in the same way as was done for spin-dependent
PDFs in Ref.~\cite{Nocera:2014uea}.

%% file: sec-exponents.tex
\section{The effective exponents}
\label{sec:exponents}

In this paper we will compute the effective exponents $\alpha_{f_i}(x,Q^2)$ and
$\beta_{f_i}(x,Q^2)$ which, when $Q^2=Q_0^2$, are asymptotically equal to the 
exponents $a_{f_i}$ and $b_{f_i}$ of the input PDF parametrisation 
Eq.~(\ref{eq:parametrisation}). Specifically, we define 
\begin{equation}
\alpha_{f_i}(x,Q^2)
\equiv
\frac{\partial\ln[xf_i(x,Q^2)]}{\partial\ln x}\, ,
\qquad
\beta_{f_i}(x,Q^2)
\equiv
\frac{\partial\ln[xf_i(x,Q^2)]}{\partial\ln(1-x)}\, ,
\label{eq:def}
\end{equation}
so that, at the input parametrisation scale $Q_0^2$, 
\begin{equation}
\alpha_{f_i}(x,Q_0^2)=
a_{f_i} 
+ x\Big[\frac{d\ln[\mathscr{F}(x,\{c_{f_i}\})]}{dx} - \frac{b_{f_i}}{1-x}\Big]
\xrightarrow{x\to 0}
a_{f_i} + O(x),
\label{eq:defasyalpha}
\end{equation}
and 
\begin{equation}
\beta_{f_i}(x,Q_0^2)=
b_{f_i} 
- (1-x)\Big[\frac{d\ln[\mathscr{F}(x,\{c_{f_i}\})]}{dx} + \frac{a_{f_i}}{x}\Big]
\xrightarrow{x\to 1}
b_{f_i} + O(1-x),
\label{eq:defasybeta}
\end{equation}
since in both Eq.~(\ref{eq:defasyalpha}) and Eq.~(\ref{eq:defasybeta}) the term 
in square brackets is by construction of order one in the corresponding limit. 
Because subasymptotic terms of $O(x)$ tend to zero very quickly at small $x$, 
and likewise subasymptotic terms of $O(1-x)$ tend to zero 
very quickly at large $x$, we expect that with the definitions 
Eq.~(\ref{eq:def}) $\alpha_{f_i}(x,Q^2)$ and $\beta_{f_i}(x,Q^2)$ 
can be used to accurately determine the asymptotic behaviour of any given PDF 
$xf_i(x,Q^2)$.
   
\begin{figure}[!t]
\centering
\includegraphics[width=0.46\textwidth,angle=270]{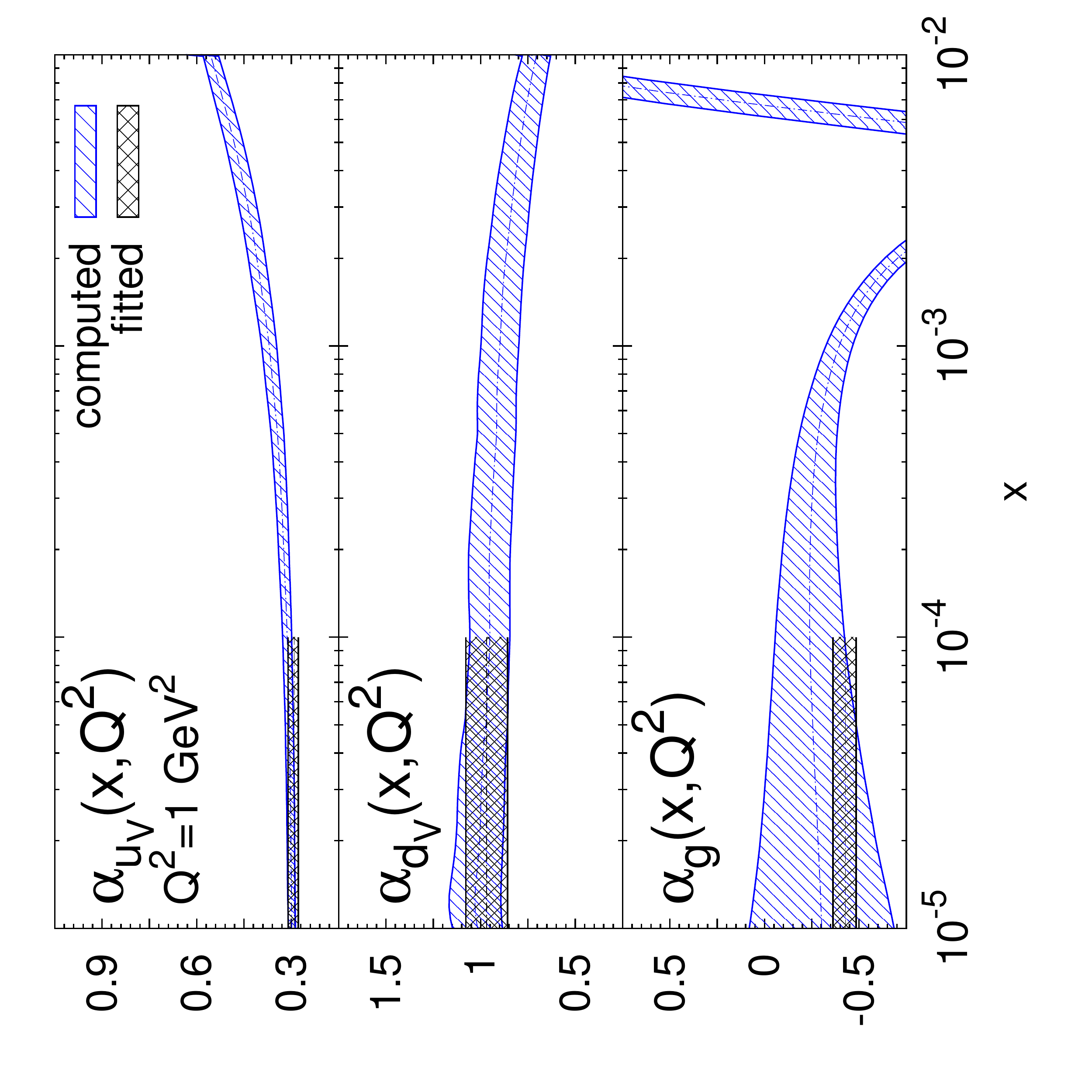}
\includegraphics[width=0.46\textwidth,angle=270]{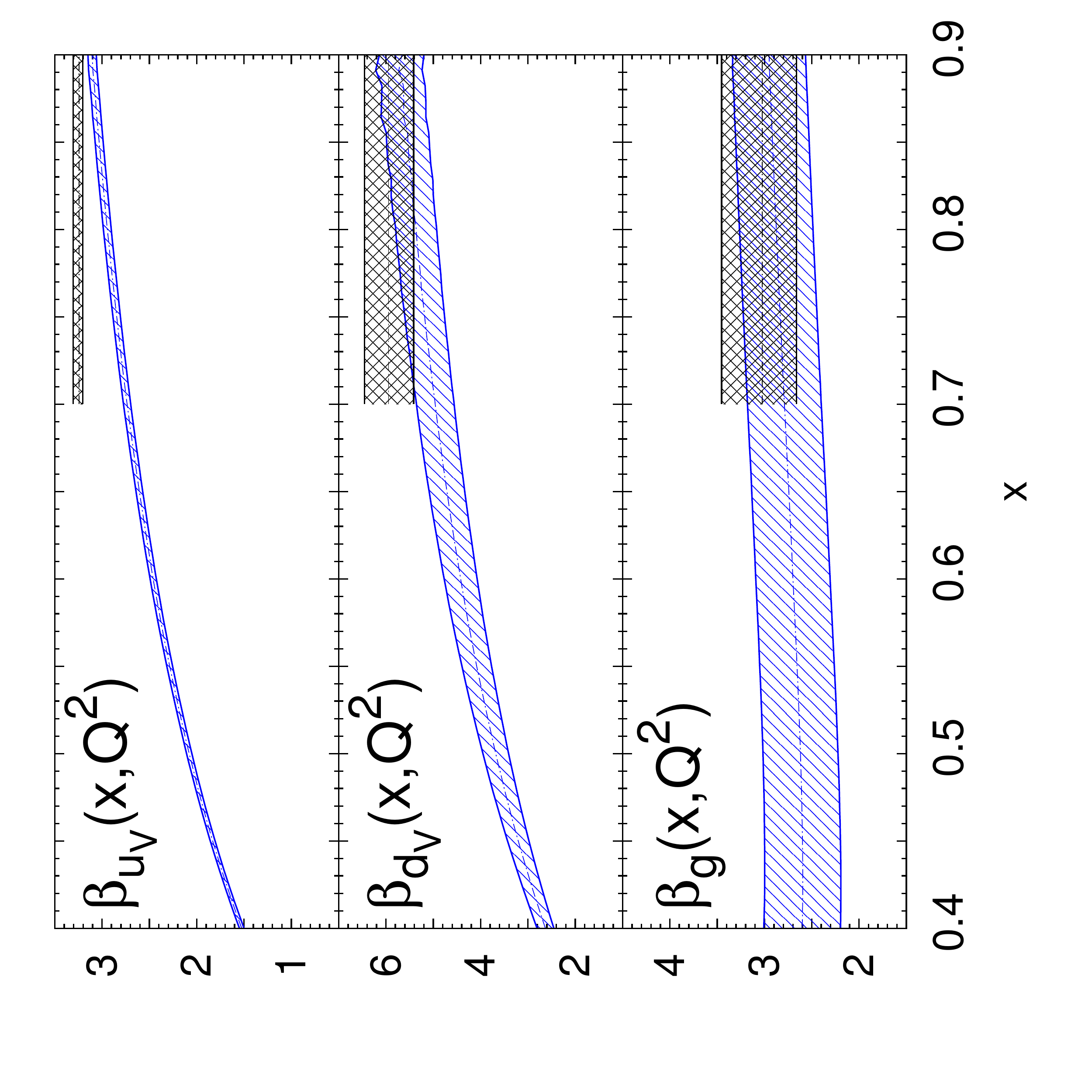}
\vspace{-0.5cm}
\caption{\small The effective exponents $\alpha_{f_i}(x,Q^2)$ 
(left) and $\beta_{f_i}(x,Q^2)$ (right) computed with Eq.~(\ref{eq:def}).
Results are shown at $Q^2=1$ GeV$^2$ for $u_V$, $d_V$ and $g$ for the  
{\tt MSTW08} NLO PDF set. 
The shaded horizontal bands indicate the fitted values of
the exponents $a_{f_i}$ (left) and $b_{f_1}$ (right) and their uncertainties.
Numerical results at $x=10^{-5}$ and $x=0.9$ are collected in
Tab.~\ref{tab:MSTW08compexp}.} 
\label{fig:testdef}
\end{figure}

In order to test this assertion, we have used Eq.~(\ref{eq:def}) to 
compute the effective asymptotic exponents $\alpha_{f_i}(x,Q^2)$
and $\beta_{f_i}(x,Q^2)$ for the {\tt MSTW08} NLO PDF set~\cite{Martin:2009iq}
(see Appendix~\ref{sec:appendix} for details).
Results at $Q^2=1$ GeV$^2$, which coincides with the input parametrisation 
scale $Q_0^2$, are shown in Fig.~\ref{fig:testdef} 
for the up valence quark, $f_i=u_V=u-\bar{u}$, the down valence quark, 
$f_i=d_V=d-\bar{d}$, and the gluon, $f_i=g$, PDFs. They are compared to the 
corresponding fitted exponents  $a_{f_i}$ and $b_{f_i}$, to which they are 
expected to approach asymptotically. 
In Tab.~\ref{tab:MSTW08compexp} we show the numerical values computed 
respectively at $x=10^{-5}$ and $x=0.9$, and again compare them
with the corresponding fitted exponents $a_i$ and $b_i$.

\begin{table}[!t]
\centering
\small
\begin{tabular}{ccccc}
\toprule
$f_{i}$ 
       & $\alpha_{f_i}(x_a,Q^2)$ 
       & $a_{f_i}$  
       & $\beta_{f_i}(x_b,Q^2)$  
       & $b_{f_i}$\\
\midrule
$u_V$ 
       & $+0.29\pm 0.01$
       & $+0.291^{+0.019}_{-0.013}$
       & $+3.11\pm 0.04$ 
       & $+3.243^{+0.062}_{-0.039}$  \\[0.2cm]
$d_V$  
       & $+1.02\pm 0.11$
       & $+0.968^{+0.110}_{-0.110}$ 
       & $+5.67\pm 0.47$ 
       & $+5.944^{+0.510}_{-0.530}$ \\[0.2cm]
$g$    
       & $-0.30\pm 0.37$
       & $-0.428^{+0.066}_{-0.057}$ 
       & $+2.95\pm 0.39$
       & $+3.023^{+0.430}_{-0.360}$ \\
\bottomrule
\end{tabular}
\normalsize
\caption{\small The effective exponents $\alpha_{f_i}$ and 
$\beta_{f_i}$ at $Q^2=1$ GeV$^2$ and $x_a=10^{-5}$ and $x_b=0.9$
computed for the  {\tt MSTW08} NLO PDF set with Eq.~(\ref{eq:def}),
compared to the corresponding fitted exponents $a_i$ and $b_i$.}
\label{tab:MSTW08compexp}
\end{table}

From Fig.~\ref{fig:testdef} and Tab.~\ref{tab:MSTW08compexp} it is
clear that both $\alpha_{f_i}(x,Q^2)$ at $x=10^{-5}$ and $\beta_{f_i}(x,Q^2)$ 
at $x=0.9$ have converged to the fitted values 
of $a_{f_i}$ and $b_{f_i}$ within PDF uncertainties.
In addition, by examining the $x$ dependence of $\alpha_{f_i}(x,Q^2)$ and 
$\beta_{f_i}(x,Q^2)$, it is possible to identify 
the asymptotic regions in which they become roughly 
independent of $x$. Furthermore, since the definitions Eq.~(\ref{eq:def}) 
may be applied at any value of $Q^2$, we may use them to study the $Q^2$ 
dependence of the effective exponents.

The definition of the PDF effective exponents, Eq.~(\ref{eq:def}), is
robust and we can therefore use it to compare the results of global fits 
among themselves and with different 
predictions from perturbative and nonperturbative QCD.
We will focus on the up and down valence PDFs,
$u_V=u-\bar{u}$ and $d_V=d-\bar{d}$, the total quark sea, 
$S=2(\bar{u} + \bar{d}) + s +\bar{s}$, and the gluon, $g$,
from the {\tt NNPDF3.0}~\cite{Ball:2014uwa}, 
{\tt MMHT14}~\cite{Harland-Lang:2014zoa}, and {\tt CT14}~\cite{Dulat:2015mca} NNLO fits.
We will also present some results from the {\tt ABM12} 
NNLO~\cite{Alekhin:2013nda} and {\tt CJ15} NLO~\cite{Accardi:2016qay} sets.
A detailed discussion of the similarities and differences between these
PDF sets can be found in Refs.~\cite{Rojo:2015acz,Ball:2015oha,Butterworth:2015oua}; here
we restrict ourselves to the information relevant for
their small and large-$x$ behaviour.
\begin{description}

\item[{\tt NNPDF3.0}]
PDFs  are parametrised in
the basis that diagonalises the DGLAP evolution equations~\cite{Ball:2008by}.
The function $\mathscr{F}(x,\{c_{f_i}\})$ is a multi-layer 
feed-forward neural network
(also known as {\it perceptron}). 
The power-law term $x^{a_{f_i}}(1-x)^{b_{f_i}}$ in Eq.~(\ref{eq:parametrisation}) 
is treated as a preprocessing factor that optimises the minimisation process:
the exponents $a_{f_i}$ and $b_{f_i}$ are chosen for each Monte Carlo replica at 
random in a given range determined iteratively.

\item[{\tt MMHT14}]
The PDFs parametrised are the valence distributions $u_V$ and $d_V$, the total 
sea $S$, the sea asymmetry $\Delta_S=\bar{d}-\bar{u}$, the total and valence 
strange distributions $s^+=s+\bar{s}$ and $s^-=s-\bar{s}$ and the gluon $g$.
The function $\mathscr{F}(x,\{c_{f_i}\})$ is taken to be a linear combination 
of Chebyshev polynomials. The exponents $a_{f_i}$ and $b_{f_i}$ are 
fitted, except for $a_{s^+}=a_{S}$. 

\item[{\tt CT14}] 
The PDFs parametrised are the
valence distributions $u_V$ and $d_V$, the sea quark distributions 
$\bar{u}$ and $\bar{d}$, the total strangeness $s^+$ and the 
gluon $g$. It is assumed that $s=\bar{s}$. 
The function $\mathscr{F}(x,\{c_{f_i}\})$ is a linear 
combination of Bernstein polynomials. The exponents $a_{f_i}$ and $b_{f_i}$ 
are parameters of the fit, but not all of them are free: specifically, it 
is assumed that $b_{u_V}=b_{d_V}$, so that as $x\to 1$ 
$u_V(x,Q_0^2)/d_V(x,Q_0^2)\to k$, with $k$ a constant, and 
that as $x\to 0$ $\bar{u}(x,Q_0^2)/\bar{d}(x,Q_0^2)\to 1$, 
which requires $a_{\bar{u}}=a_{\bar{d}}$.

\item[{\tt ABM12}]
The PDFs parametrised are the valence distributions $u_V$ and $d_V$, the sea 
distributions $\bar{u}$ and $s$, the sea asymmetry $\Delta_S$ and the gluon 
$g$. It is assumed that $s=\bar{s}$. The function $\mathscr{F}(x,\{c_{f_i}\})$ 
has the form $x^{P_{f_i}(x)}$, where $P_{f_i}(x)$ is a function of $x$; for $s$, 
$\mathscr{F}(x,\{c_{f_i}\}=1$. The exponents $a_{f_i}$ and $b_{f_i}$ are parameters
of the fit, except for the condition $a_{\Delta_S}=0.7$.

\item[{\tt CJ15}]
The PDFs parametrised are the valence distributions $u_V$ and $d_V$, 
the light antiquark sea, $\bar{u}+\bar{d}$, the light antiquark ratio 
$\bar{d}/\bar{u}$, the total strangeness $s^+$ and the gluon $g$. It is assumed 
that $s=\bar{s}$. The function $\mathscr{F}(x,\{ c_{f_i}\})$ is provided by the 
polynomial $(1+c_{f_i}^{(1)}\sqrt{x}+c_{f_i}^{(2)}x)$ for all the 
distributions except the light antiquark ratio and the total strangeness. 
Specifically, $\bar{d}/\bar{u}$ is parametrised with a simple polynomial
which ensures that as $x\to 1$, $\bar{d}/\bar{u}\to 1$, while it is assumed 
that $s^+=\kappa(\bar{u}+\bar{d})$; $c_{f_i}^{(1)}$, $c_{f_i}^{(2)}$ and
$\kappa$ are parameters of the fit. A small admixture of $u_V$ is added to
$d_V$ so that as $x\to 1$ $d_V/u_V\to k$, with $k$ a constant.

\end{description}

Although the momentum distributions of strange and antistrange quarks
are assumed to be identical in some of these PDF sets, it should be noted that 
a strange/antistrange asymmetry in the nucleon is predicted based on 
nonperturbative QCD models, see {\it e.g.} Ref.~\cite{Brodsky:1996hc} and
references therein. Strange and antistrange distributions may also
be very different with each other in the polarized case, as it was shown in 
Ref.~\cite{Brodsky:1996hc} based on a light-cone model of 
energetically-favoured meson-baryon fluctuations applied to the 
$K^+\Lambda$. However, a study of a structured asymmetry in the momentum 
distributions of strange and antistrange quarks in a global QCD analysis 
is beyond the scope of this work, and has been addressed 
elsewhere~\cite{Ball:2014uwa,Harland-Lang:2014zoa}.

\begin{figure}[!t]
\centering
\includegraphics[width=0.46\textwidth,angle=270]{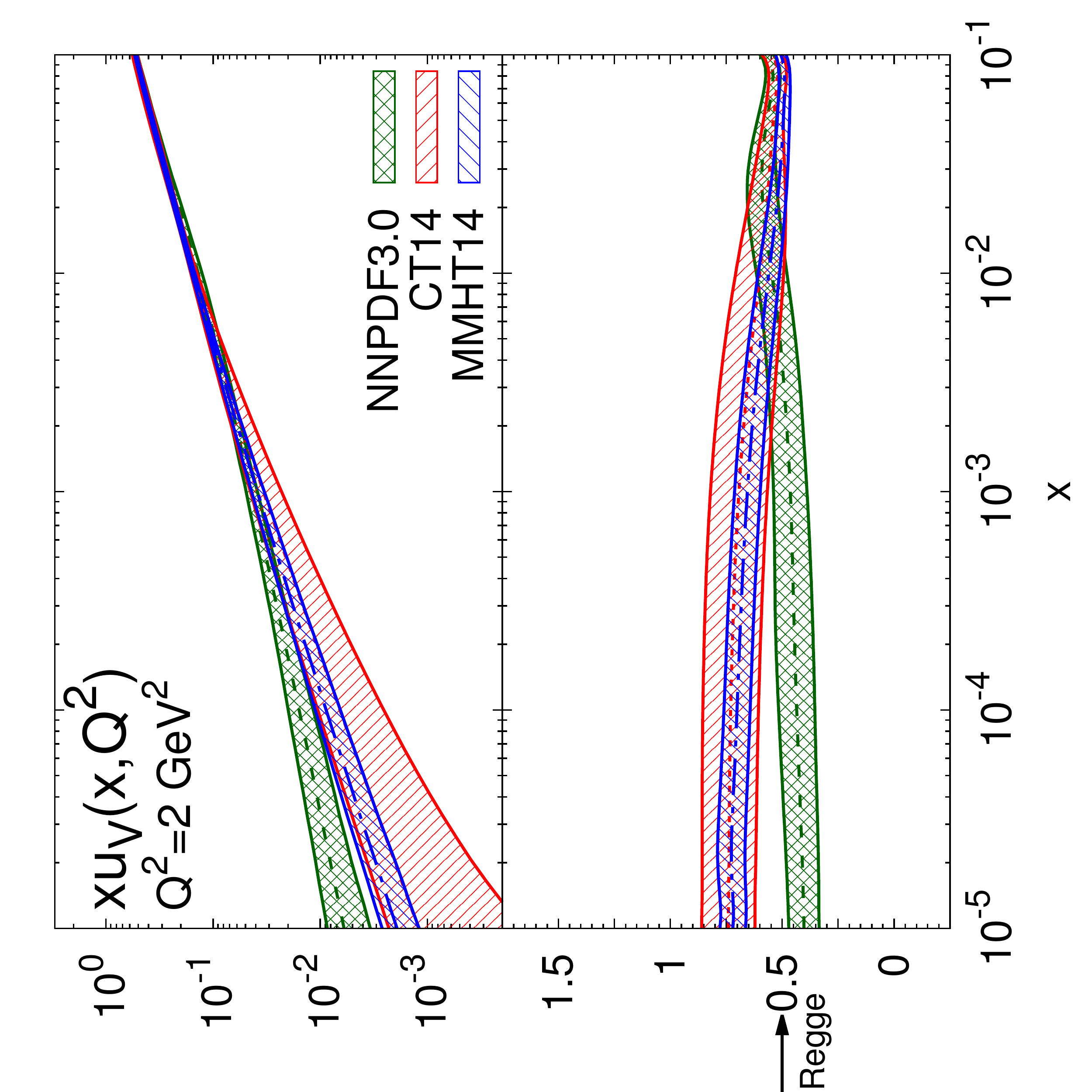}
\includegraphics[width=0.46\textwidth,angle=270]{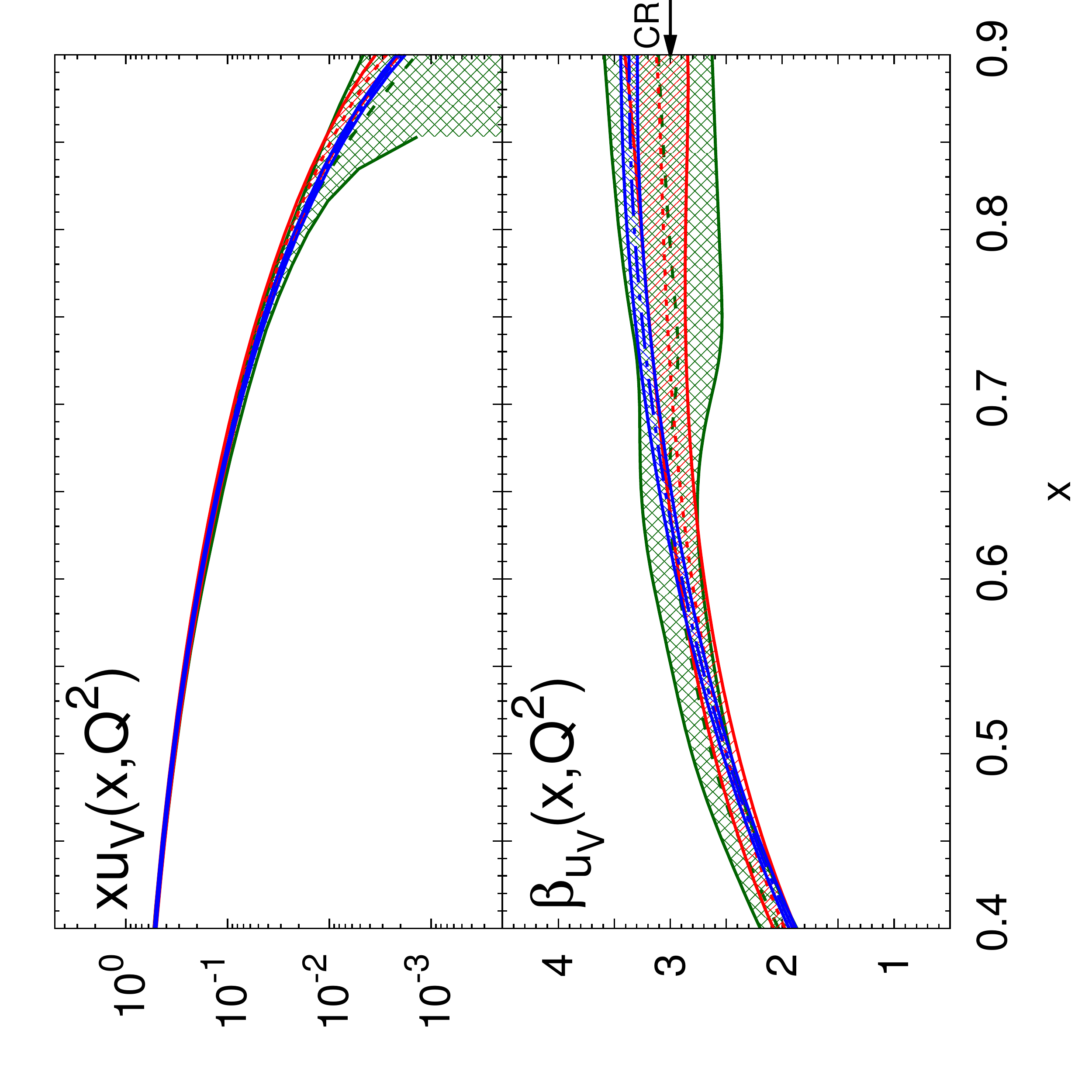}\\
\includegraphics[width=0.46\textwidth,angle=270]{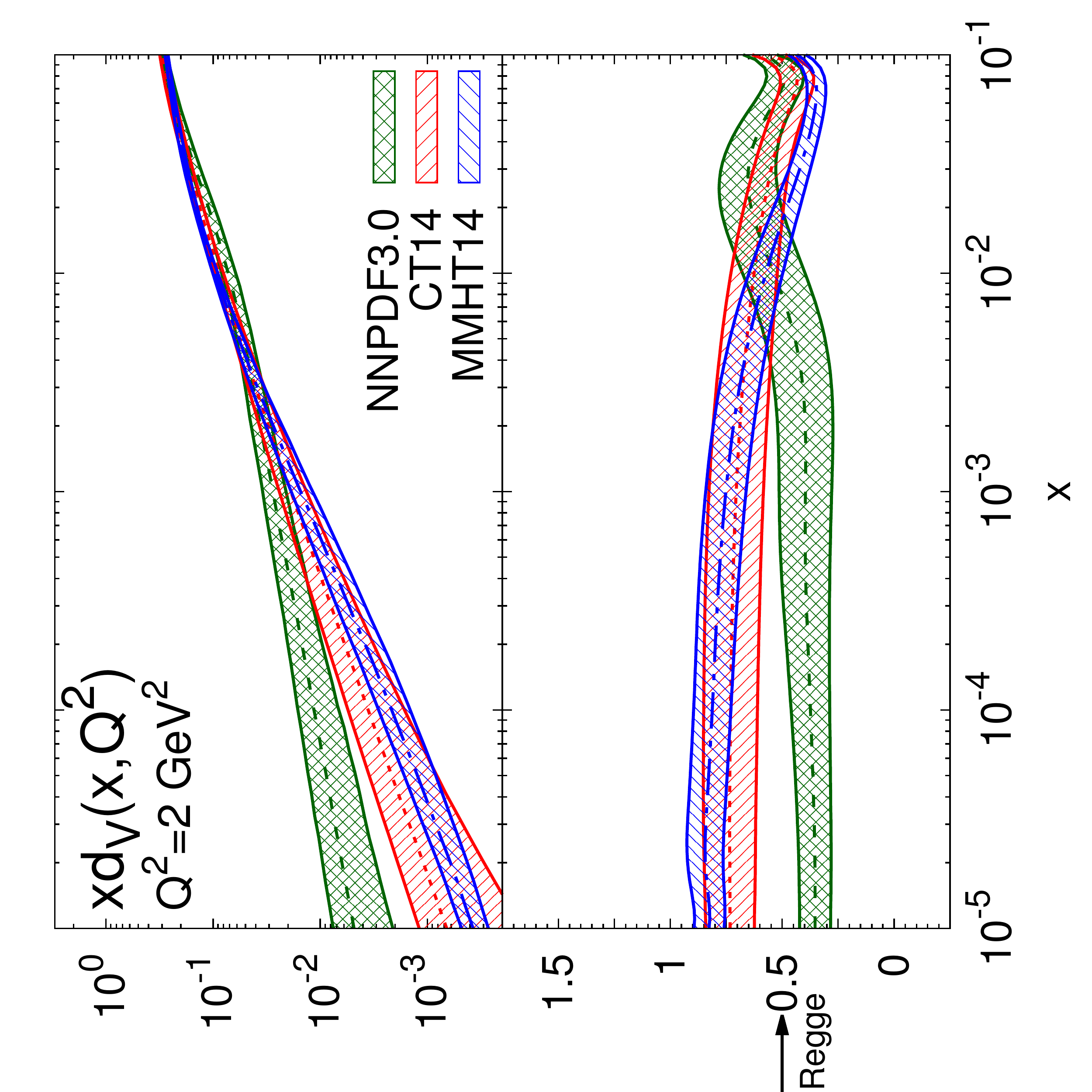}
\includegraphics[width=0.46\textwidth,angle=270]{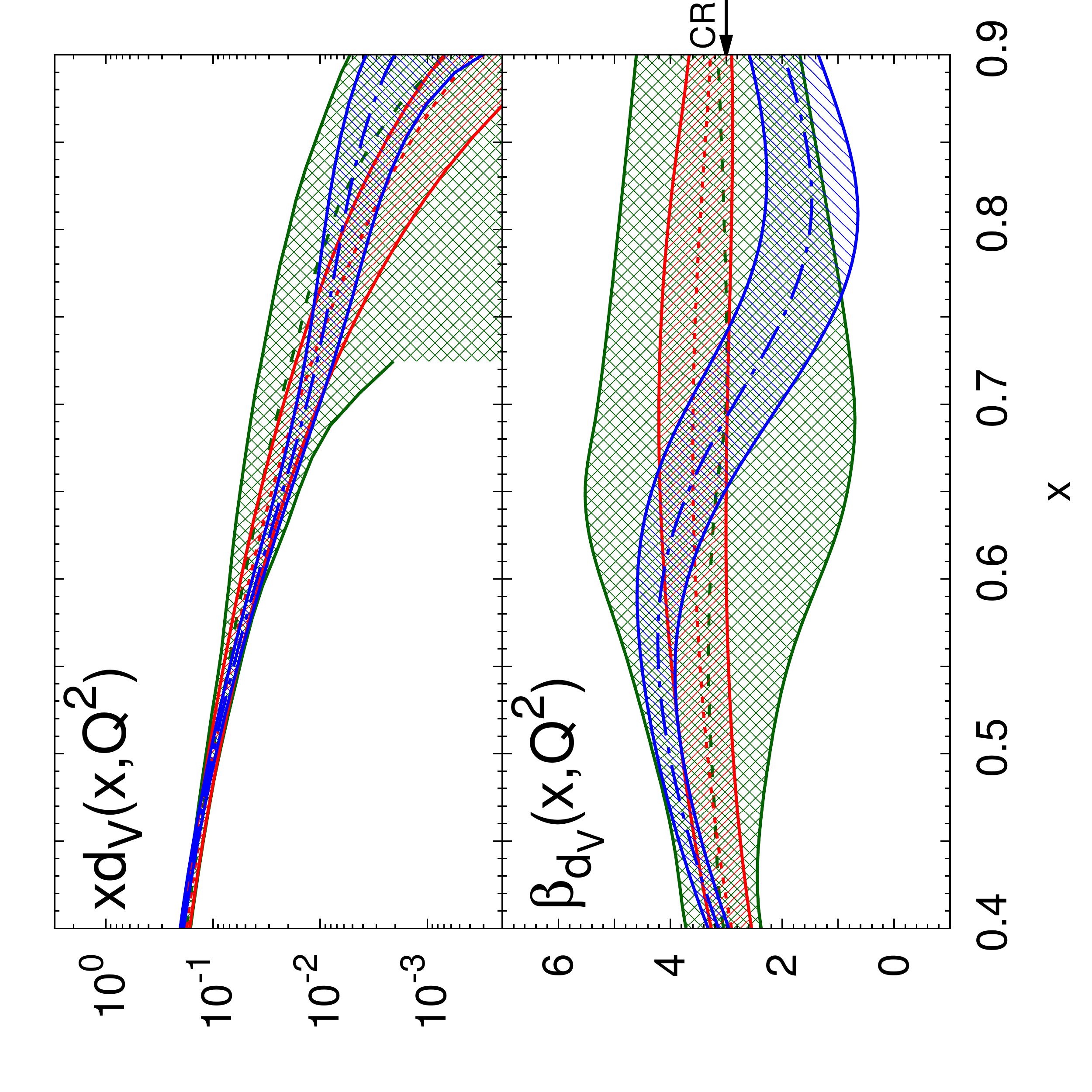}\\
\vspace{-0.3cm}
\caption{\small The effective exponents $\alpha_{f_i}(x,Q^2)$ (left) and 
$\beta_{f_i}(x,Q^2)$ (right), Eq.~(\ref{eq:def}), for the up 
valence (top) and down valence (bottom) PDFs, as a function of $x$ 
at $Q^2=2$ GeV$^2$, together with the corresponding PDFs.
Results are shown for the {\tt NNPDF3.0}, {\tt CT14} and {\tt MMHT14} NNLO 
PDF sets. The arrows indicate the prediction from Regge theory (Regge) and 
Brodsky-Farrar quark counting rules (CR).}
\label{fig:expeffval}
\end{figure}

\begin{figure}[!t]
\centering
\includegraphics[width=0.46\textwidth,angle=270]{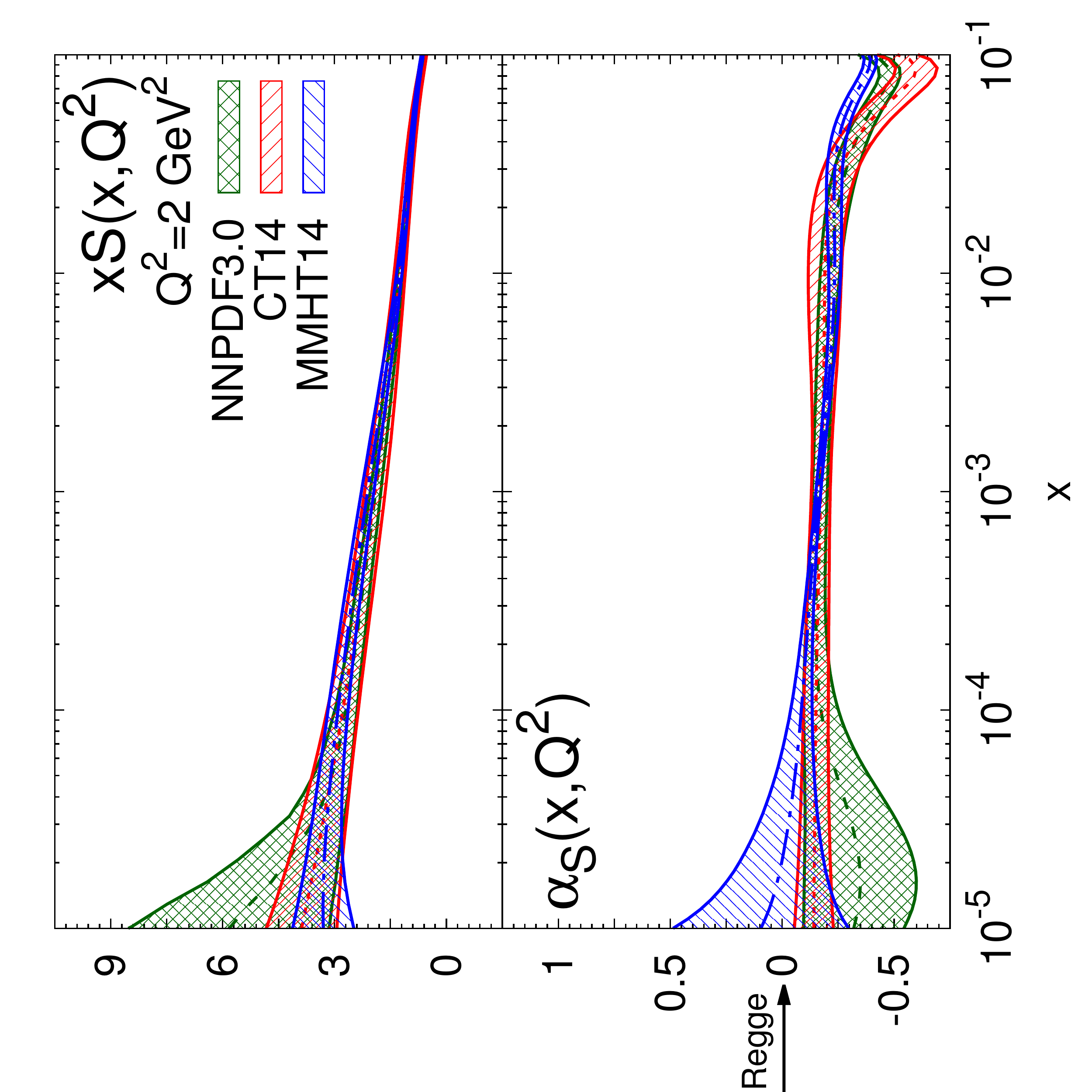}
\includegraphics[width=0.46\textwidth,angle=270]{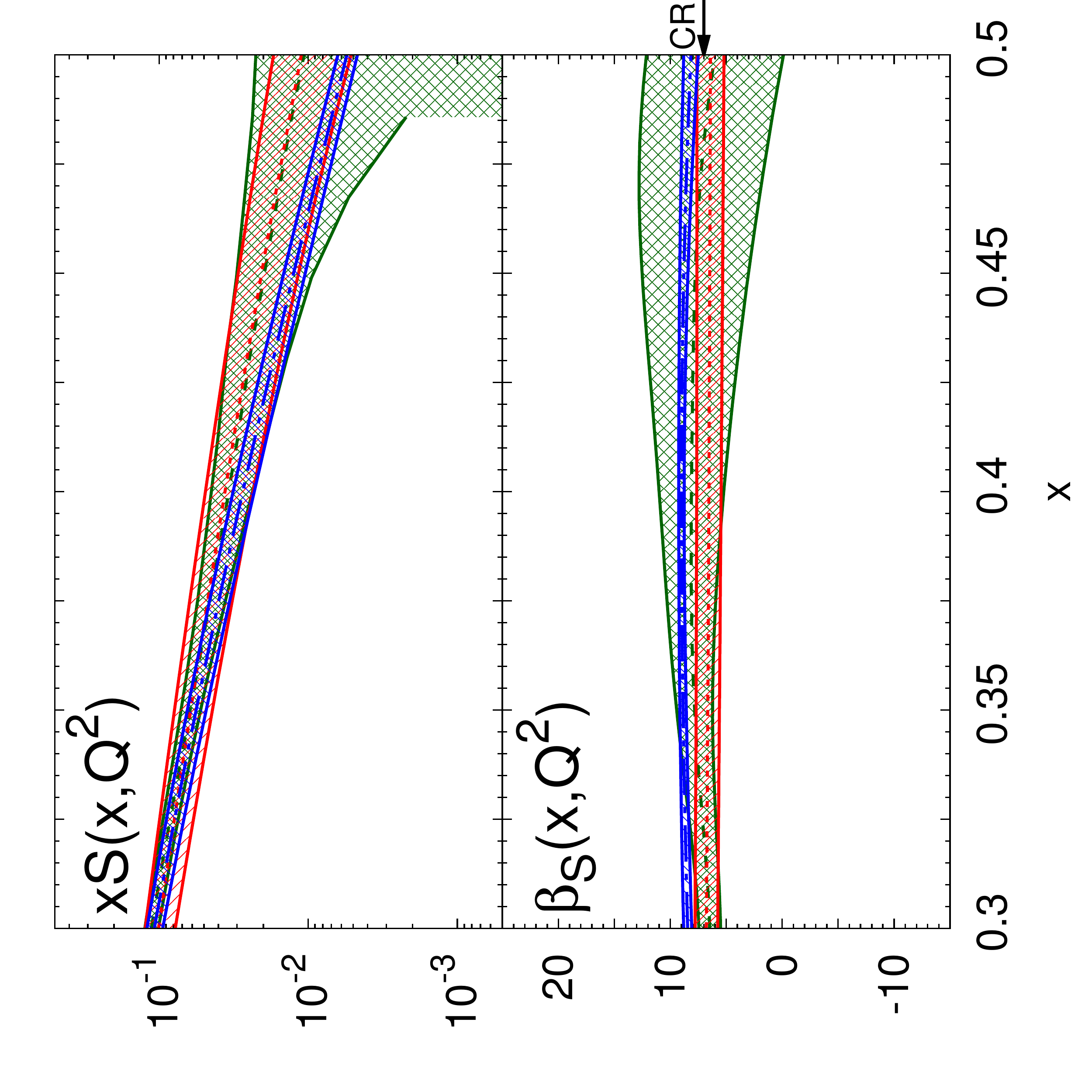}
\vspace{-0.3cm}
\caption{\small Same as Fig.~\ref{fig:expeffval} for the sea PDF 
$S(x,Q^2)$.}
\label{fig:expeffseadelta}
\end{figure}

\begin{figure}[!t]
\centering
\includegraphics[width=0.46\textwidth,angle=270]{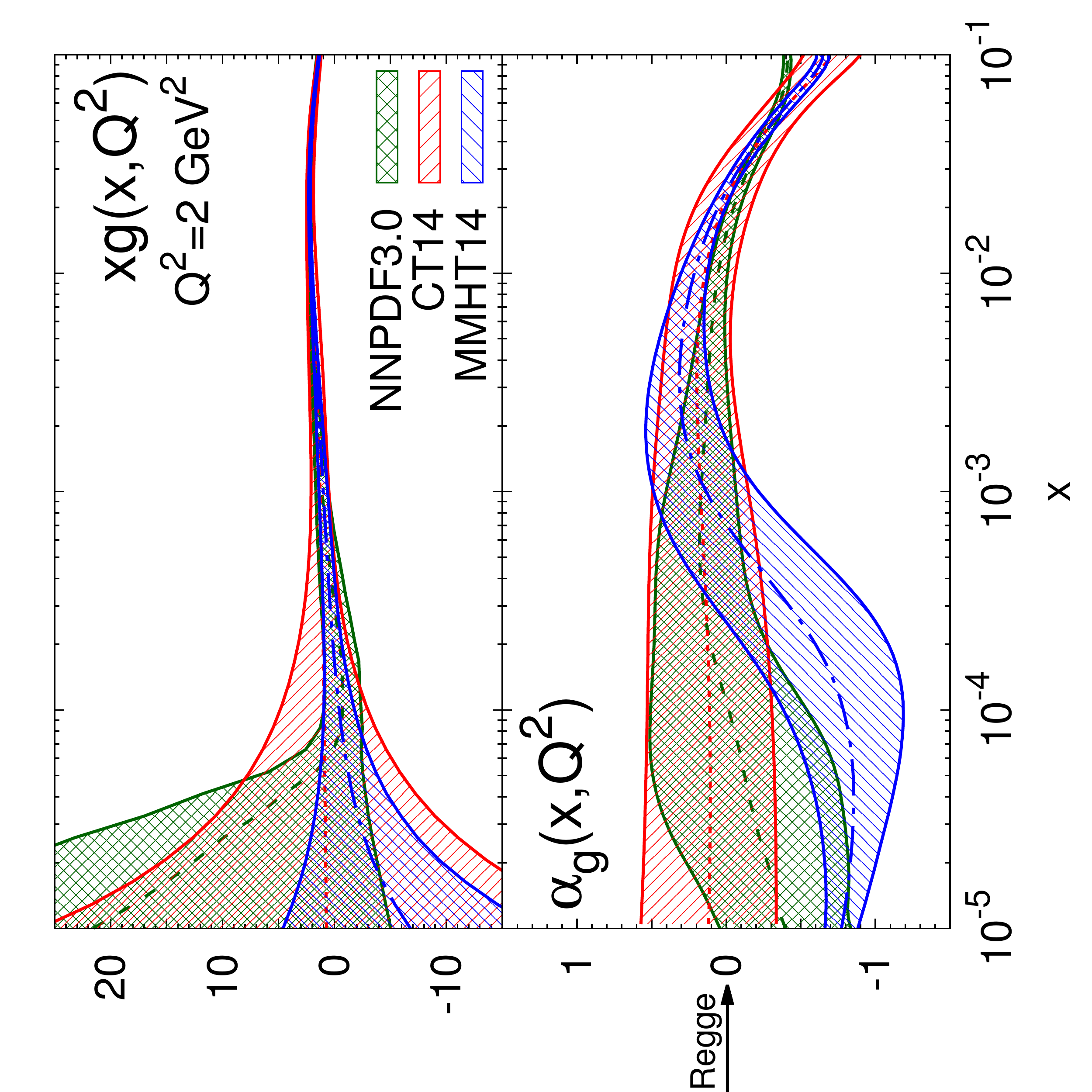}
\includegraphics[width=0.46\textwidth,angle=270]{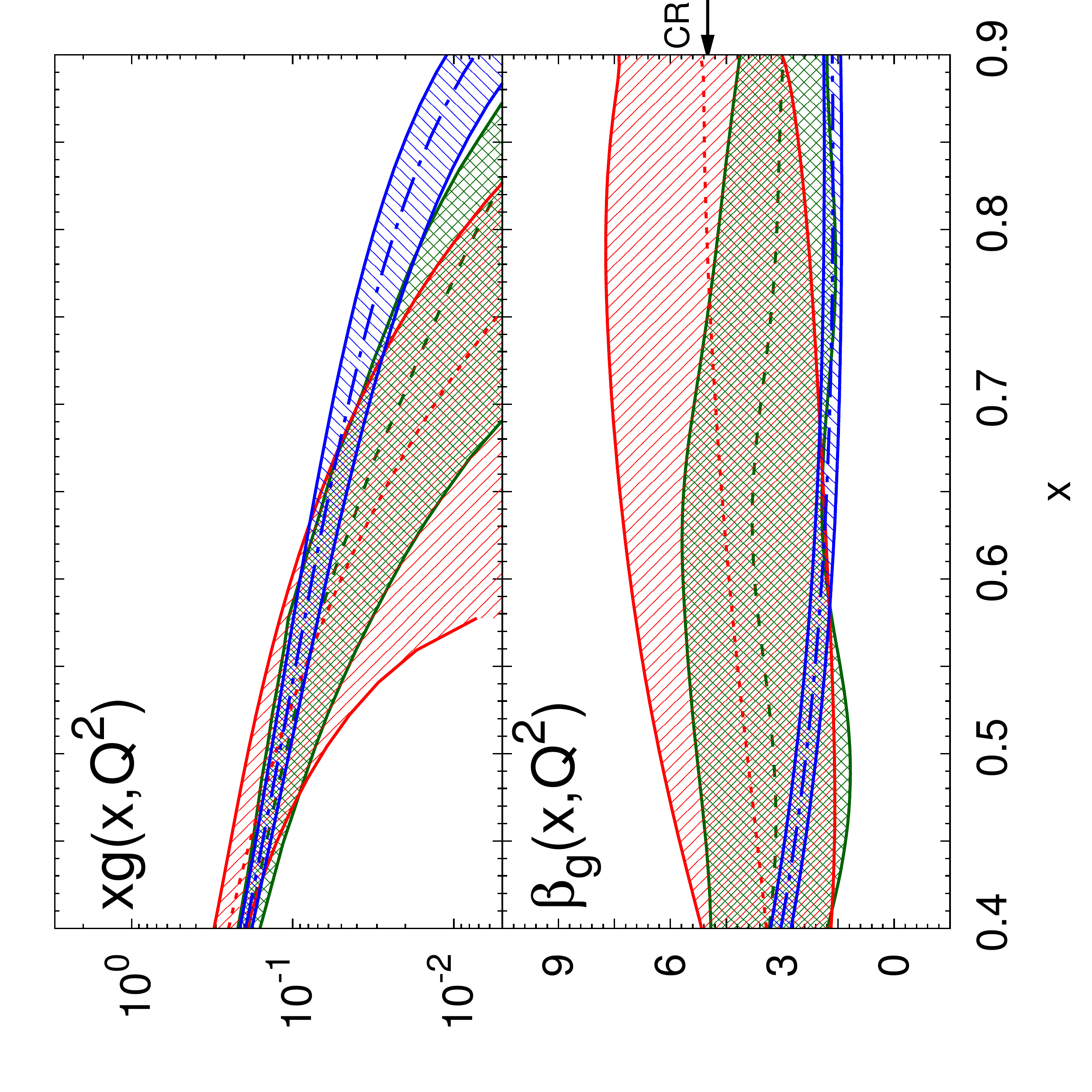}
\vspace{-0.3cm}
\caption{\small Same as Fig.~\ref{fig:expeffval} for the gluon PDF 
$g(x,Q^2)$.}
\label{fig:expeffgluon}
\end{figure}

In Figs.~\ref{fig:expeffval}-\ref{fig:expeffgluon} we compare both the PDFs 
and the corresponding effective exponents $\alpha_{f_i}(x,Q^2)$ and 
$\beta_{f_i}(x,Q^2)$ for the {\tt NNPDF3.0}, {\tt MMHT14} and {\tt CT14}
sets at $Q^2=2$ GeV$^2$. For {\tt NNPDF3.0}, PDF uncertainties
are computed as $68\%$ confidence level (CL) intervals, while for
{\tt MMHT14} and {\tt CT14} sets we show the symmetric 
one-sigma Hessian uncertainties.
In most cases it is possible to identify an asymptotic region where the 
effective exponents become approximately independent of $x$. The onset of this 
asymptotic regime depends on both the PDF flavour and on the PDF set.
At small $x$, the asymptotic regime is reached at $x\lesssim 10^{-3}$ 
for $u_V$, $d_V$ and $S$ irrespective of the PDF set considered.
For the gluon, convergence is achieved at smaller values of $x$, $x\lesssim 10^{-5}$, 
at least for {\tt MMHT14} for which $\alpha_g(x,Q^2)$ has an oscillation
in the region  $10^{-4}\lesssim x\lesssim 10^{-3}$.
Note that at $x\lesssim 10^{-4}$ PDFs are extrapolated into a region
with very limited experimental information. This very small-$x$ region can be 
probed at the LHC with forward charm~\cite{Gauld:2015yia,Zenaiev:2015rfa} and 
quarkonium production~\cite{Jones:2015nna}. 
At large $x$, the asymptotic region is 
reached at $x\gtrsim 0.7$ in most cases.
The exception is $\beta_{d_V}(x,Q^2)$ from {\tt MMHT14}, which exhibits an  
oscillation in the region $0.6\lesssim x\lesssim 0.8$. 

To the best of our knowledge, this is the first time that the onset of an 
asymptotic regime in the effective PDF exponents $\alpha_{f_i}(x,Q^2)$ and 
$\beta_{f_i}(x,Q^2)$ has been explicitly demonstrated. Remarkably, this onset 
takes place at $x$ values close to the boundary between the data and 
extrapolation regions.
Our results indicate that the three global PDF sets are broadly 
consistent among one other within uncertainties not only at the level of PDFs, 
but also at the level 
of their small- and large-$x$ asymptotic behaviour.
The main exceptions are $u_V$ and $d_V$ at small $x$, where the effective
exponent of {\tt NNPDF3.0} is incompatible with those of {\tt CT14} and 
{\tt MMHT14}. However, this is an extrapolation region where the Hessian 
approximation has some limitations and non-Gaussian effects are large: indeed,
if we compute with {\tt NNPDF3.0} the one-sigma PDF interval as opposed to 
the $68\%$ CL, the three sets become consistent.

\begin{figure}[!t]
\centering
\includegraphics[width=0.46\textwidth,angle=270]{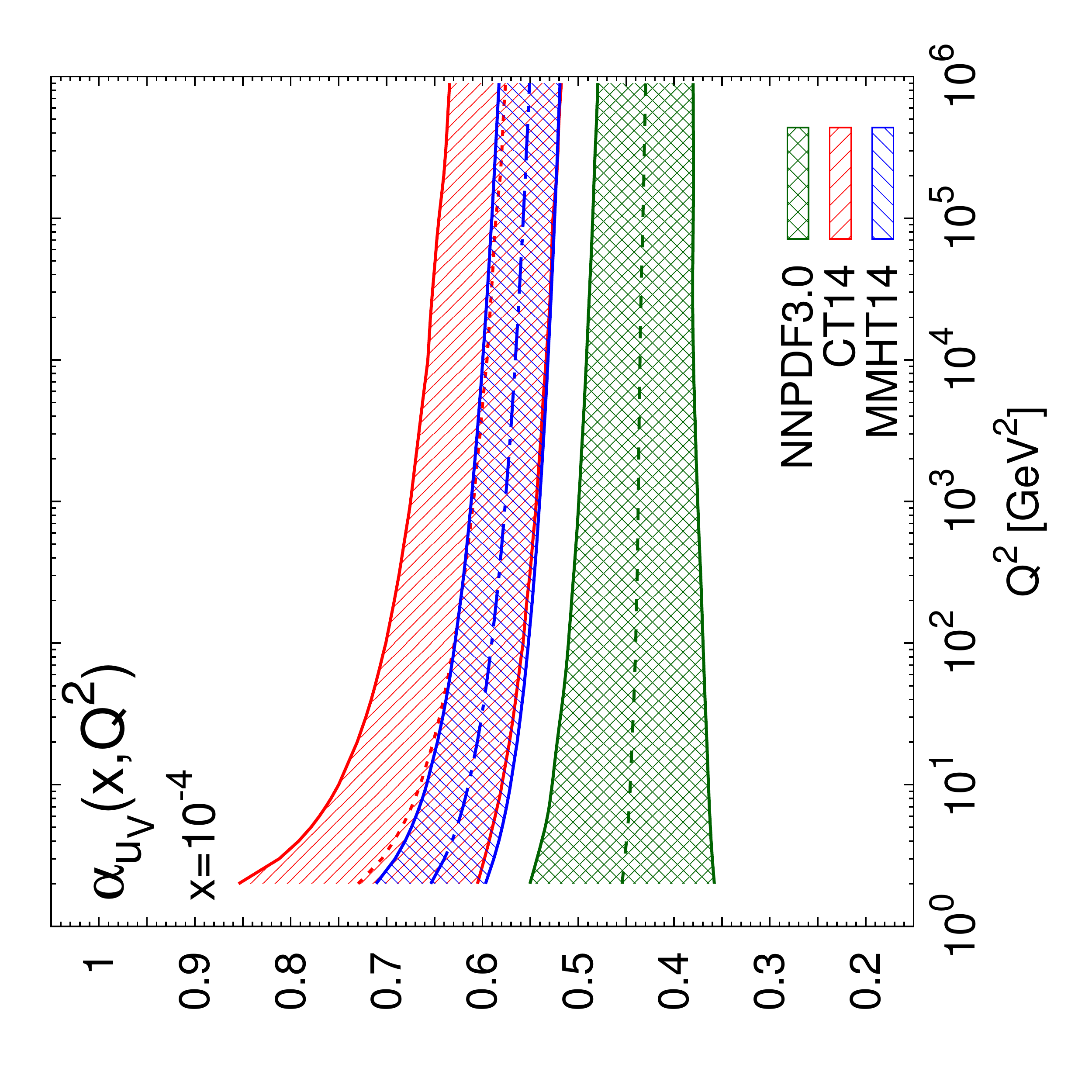}
\includegraphics[width=0.46\textwidth,angle=270]{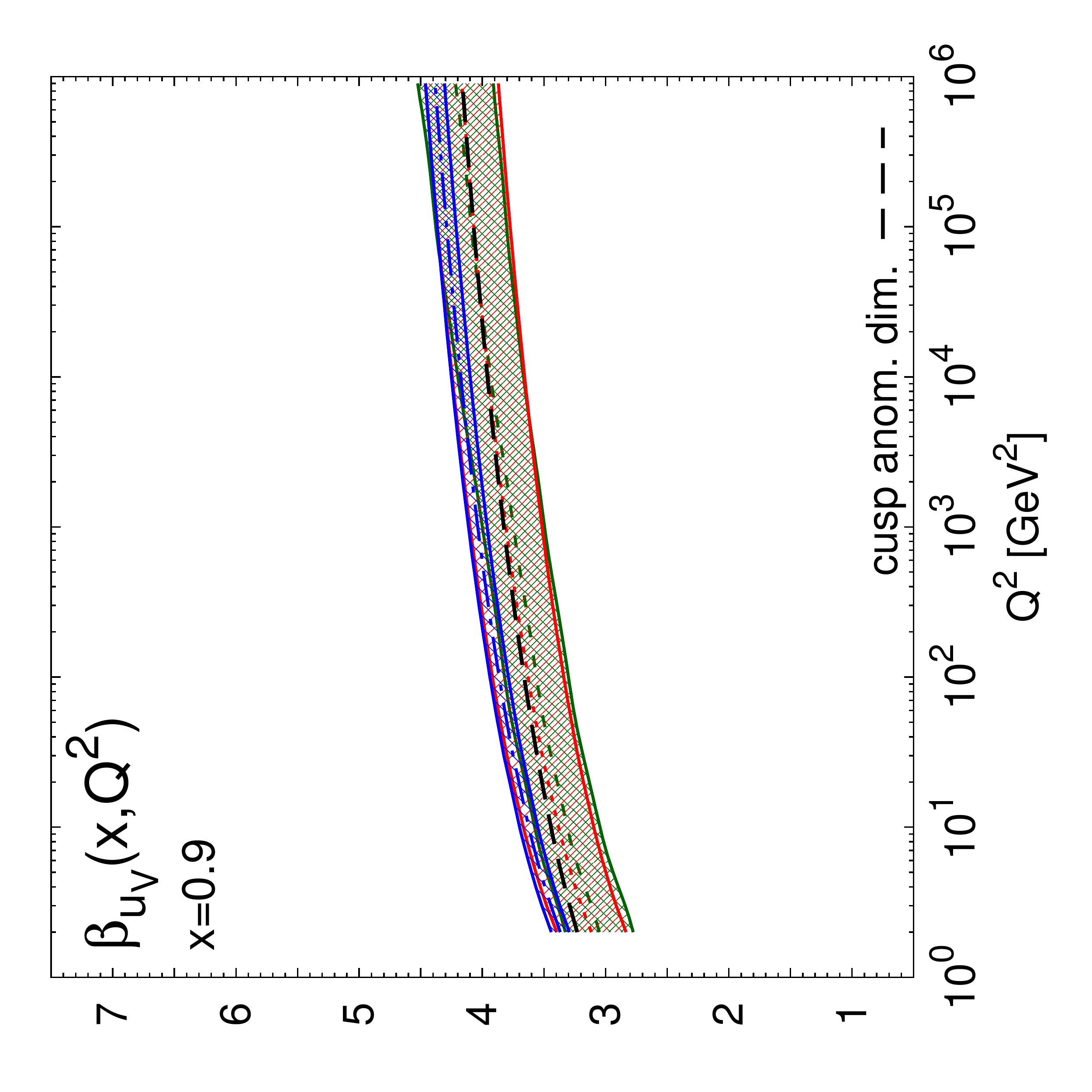}\\
\includegraphics[width=0.46\textwidth,angle=270]{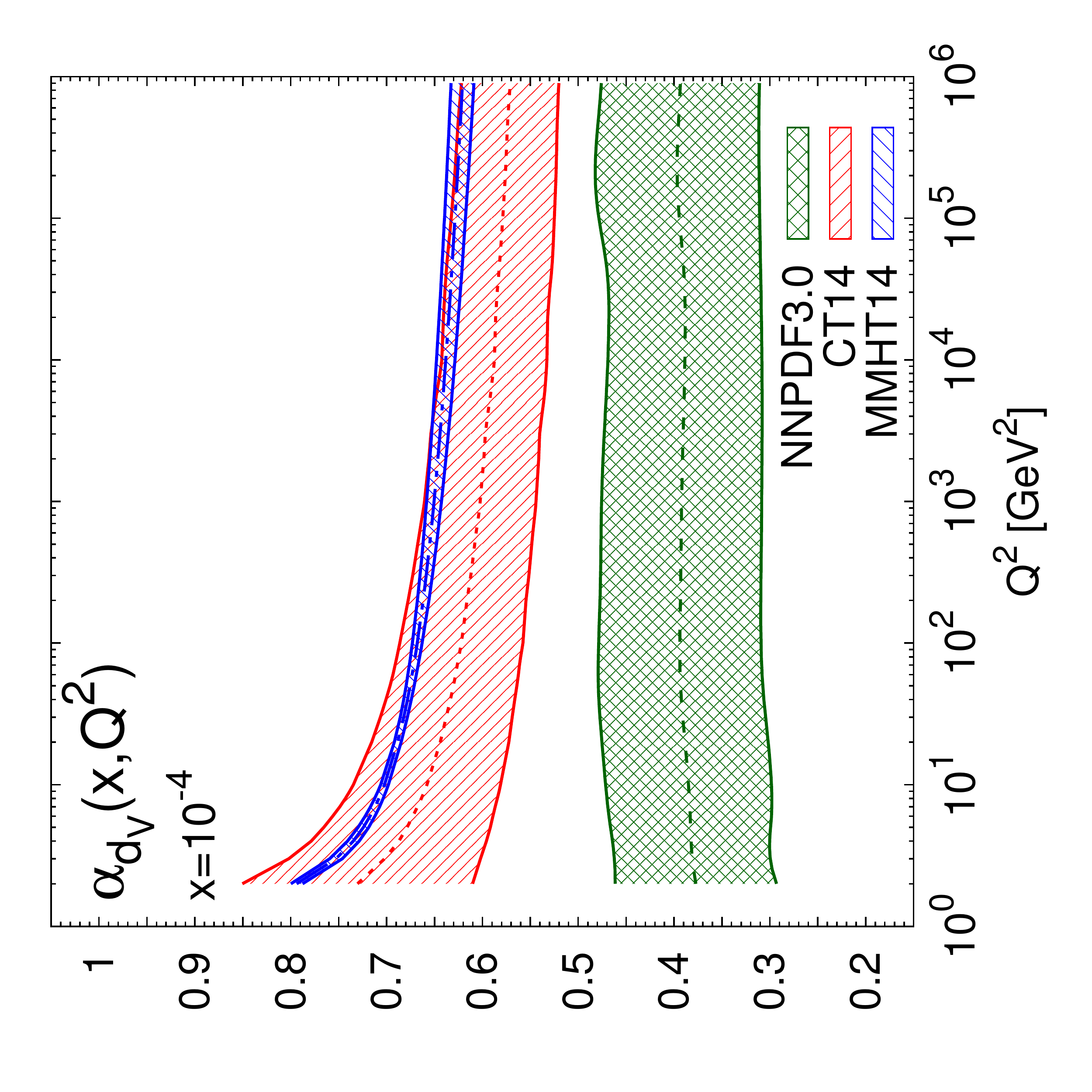}
\includegraphics[width=0.46\textwidth,angle=270]{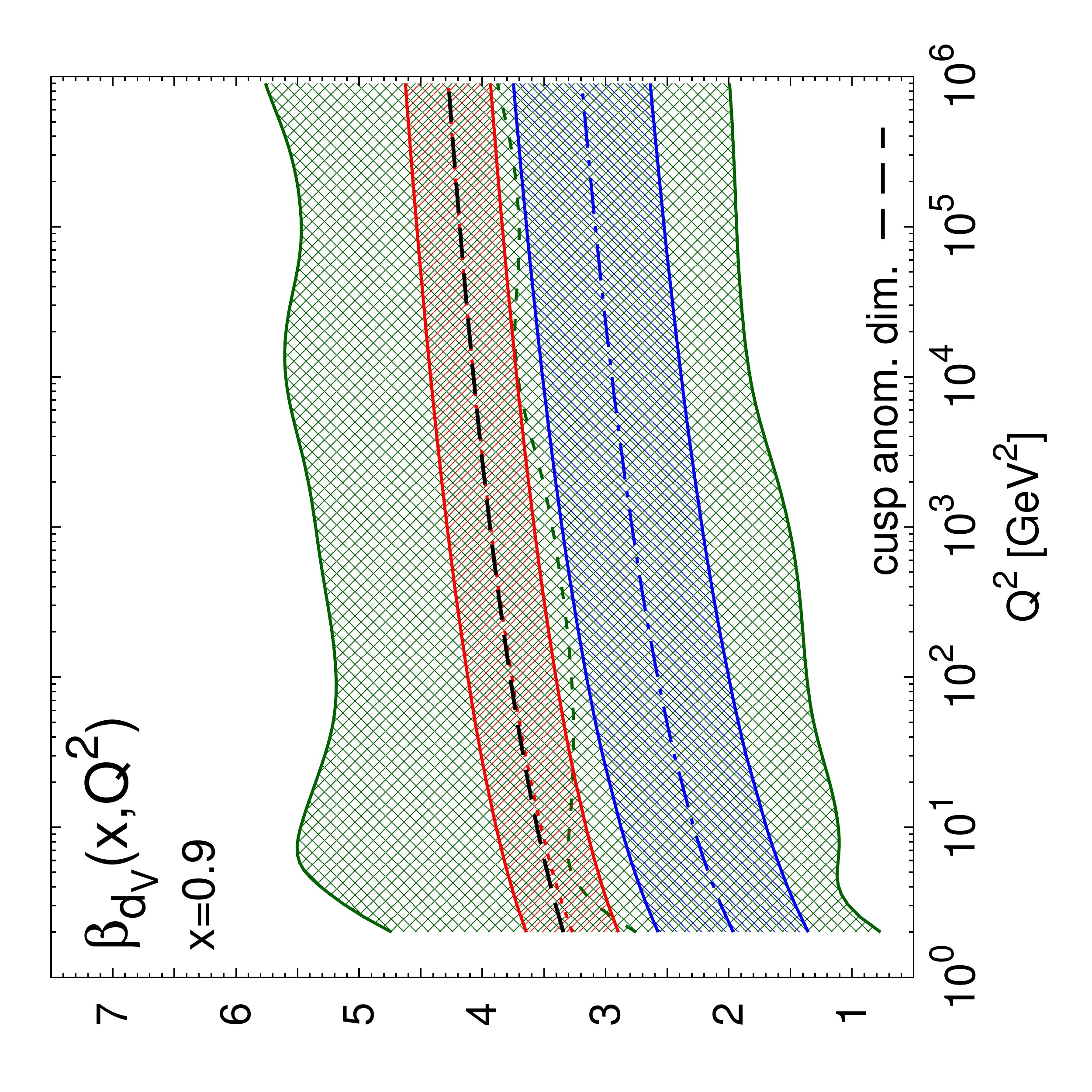}
\vspace{-0.3cm}
\caption{\small The effective exponents $\alpha_{f_i}(x,Q^2)$ (left) and 
$\beta_{f_i}(x,Q^2)$ (right), Eq.~(\ref{eq:def}), for the up (top) 
and down valence (bottom) PDFs, as a function of $Q^2$ at $x=10^{-4}$ and 
$x=0.9$ respectively, for the {\tt NNPDF3.0}, {\tt CT14} and
{\tt MMHT14} NNLO sets. At large $x$, the perturbative QCD prediction
Eq.~(\ref{eq:betascale}) is also
displayed for {\tt CT14}.}
\label{fig:evolvalence}
\end{figure}

\begin{figure}[!t]
\centering
\includegraphics[width=0.46\textwidth,angle=270]{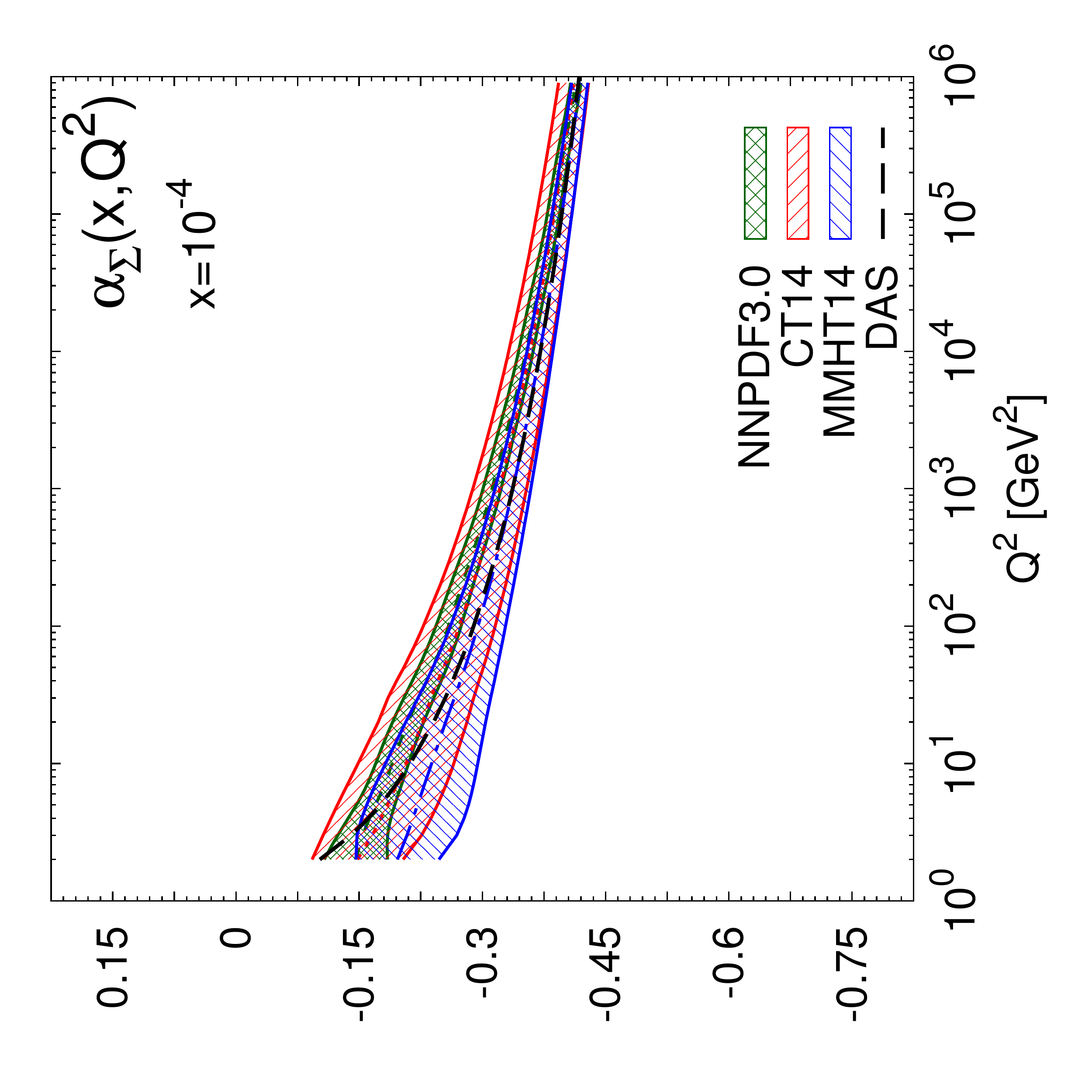}
\includegraphics[width=0.46\textwidth,angle=270]{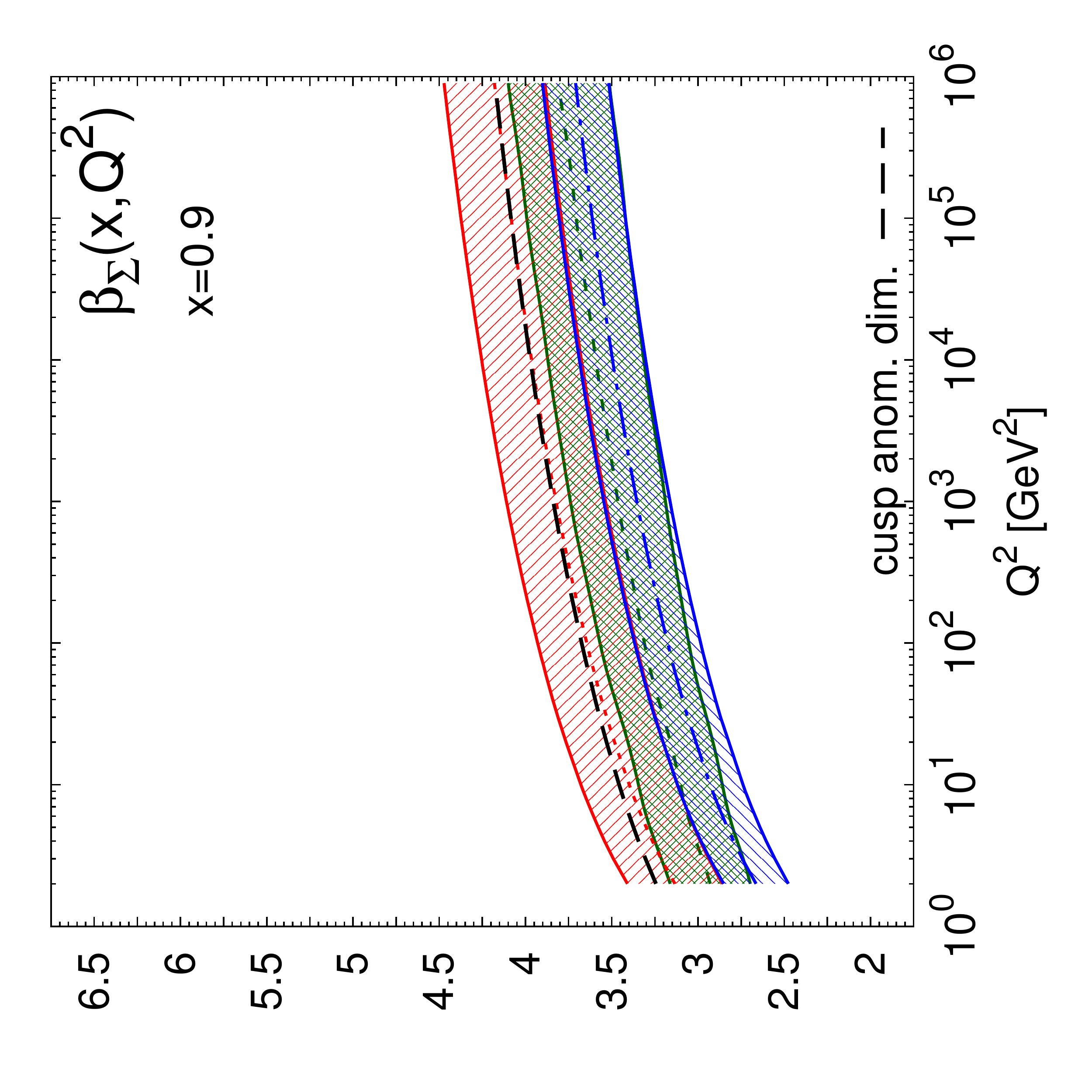}\\
\includegraphics[width=0.46\textwidth,angle=270]{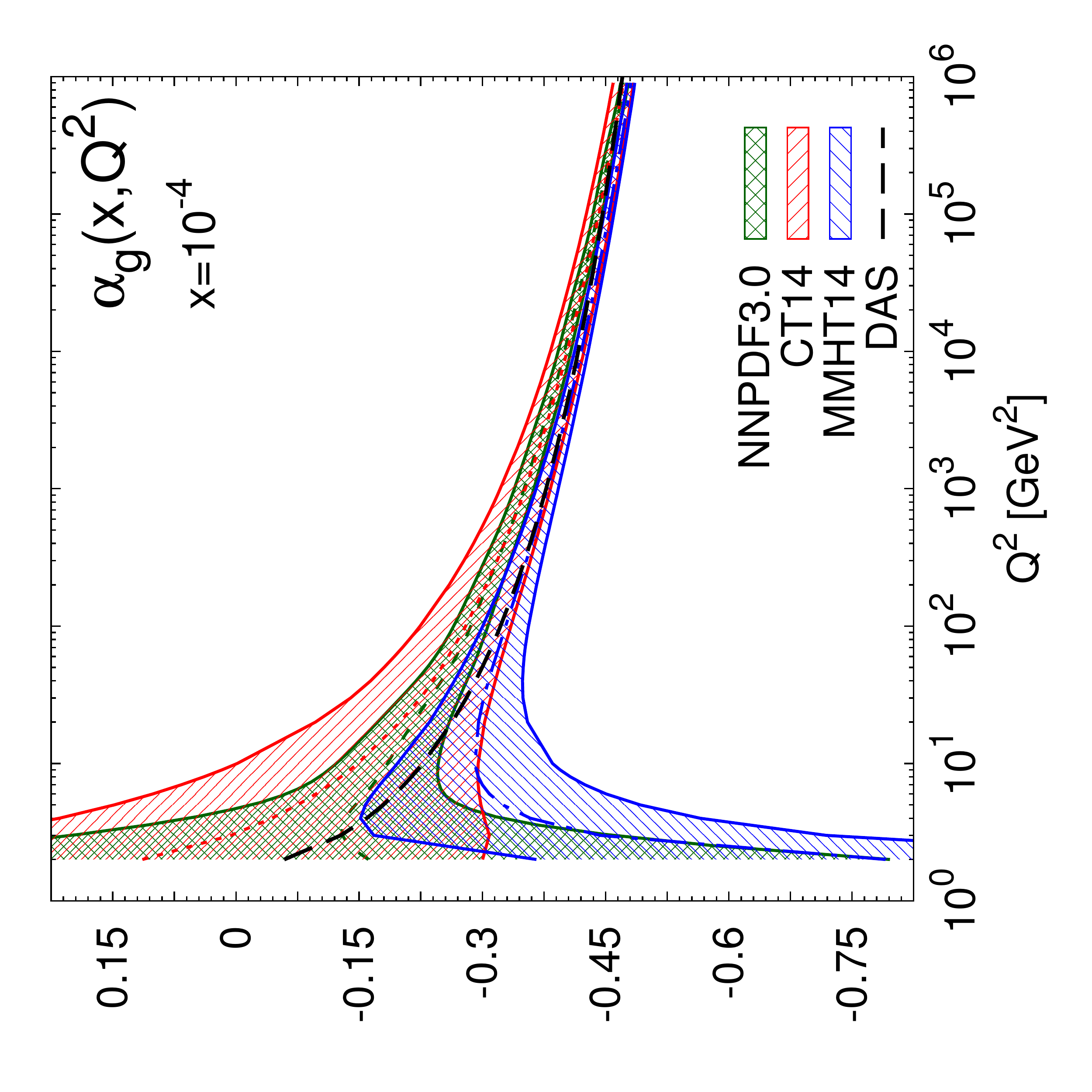}
\includegraphics[width=0.46\textwidth,angle=270]{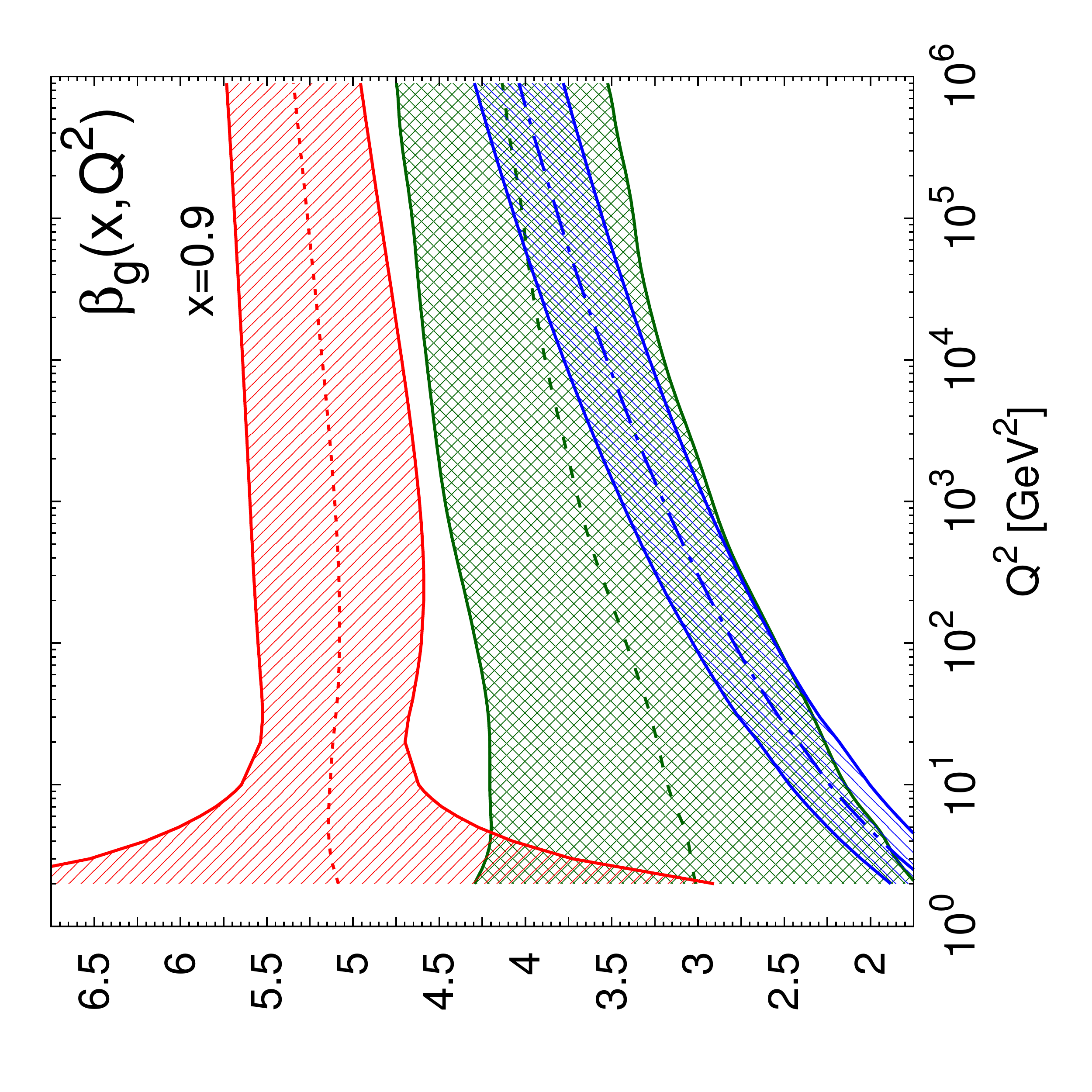}
\vspace{-0.3cm}
\caption{\small Same as Fig.~\ref{fig:evolvalence} for the quark singlet 
  $\Sigma$ and the gluon $g$. For $x=10^{-4}$ (left plots) we also show the DAS predictions,
  Eq.~(\ref{eq:dasvalues}).}
\label{fig:evolsinglet}
\end{figure}

Before we compare our results to the expectations of Regge theory and the 
Brodsky-Farrar quark counting rules, we first examine the $Q^2$ dependence of 
the effective exponents. 
To this end, in Figs.~\ref{fig:evolvalence}-\ref{fig:evolsinglet} we show 
the effective exponents $\alpha_{f_i}(x,Q^2)$ and $\beta_{f_i}(x,Q^2)$ as 
functions of $Q^2$ at fixed values of $x$ in the asymptotic region: 
$x=10^{-4}$ and $x=0.9$ respectively.
We show results for the valence distributions $u_V$ and $d_V$, the total 
quark singlet $\Sigma=\sum_{i=1}^{n_f} (q_i + \bar{q}_i)$ and the gluon.
From these plots we can see that as $Q^2$ 
increases the effective exponents become less sensitive to $Q^2$ and tend to 
converge to a finite
value asymptotically. This feature is broadly independent of $x$ when $x$ is
sufficiently small or large, roughly $x\lesssim 10^{-3}$ and $x\gtrsim 0.9$.
The only exception is again $\beta_{d_V}(x,Q^2)$ for {\tt MMHT14}.

At small $x$, the $Q^2$ dependence of the effective exponents illustrates 
the transition from a low-$Q^2$ region, where PDFs are determined from 
nonperturbative dynamics, to a high-$Q^2$ region, where PDFs are dominated 
by perturbative QCD evolution. Indeed, as $x\to 0$ and $Q^2\to \infty$, 
PDFs can be solely determined by DGLAP 
equations~\cite{Dokshitzer:1977sg,DeRujula:1974mnv},
provided that their behaviour is sufficiently soft at the input scale.
In this limit, it is known that PDFs exhibit a 
Double Asymptotic Scaling 
(DAS)~\cite{Ball:1994du,Ball:1994kc,Forte:1995vs,Ralston:1986hr}. 
Specifically,
as $x\to 0$ and $Q^2\to\infty$ the singlet sector grows as
\begin{equation}
x\Sigma(x,Q^2)\to
{\cal N}_\Sigma\frac{\gamma}{\rho}\frac{1}{\sqrt{4\pi\gamma\sigma}}
e^{2\gamma\sigma - \delta\sigma/\rho}\, ,
\qquad
xg(x,Q^2) 
\to {\cal N}_g\frac{1}{\sqrt{4\pi\gamma\sigma}}e^{2\gamma\sigma - \delta\sigma/\rho}\, ,
\label{eq:das}
\end{equation}  
where we have defined 
\begin{equation}
\gamma \equiv \left(\frac{12}{\beta_0} \right)^{1/2}\, ,\qquad 
\delta \equiv \left(11 + \frac{2n_f}{27}\right)\Big/\beta_0,\qquad
\beta_0 = 11 - \frac{2}{3}n_f\, ,
\label{eq:das1}
\end{equation}
and the double scaling variables
\begin{equation}
\sigma 
\equiv
\left[
\ln\frac{x_0}{x}
\ln\frac{\ln\left(Q^2/\Lambda^2\right)}{\ln\left(Q_0^2/\Lambda^2\right)}
\right]^{1/2},\qquad
\rho
\equiv
\left[
\frac{\ln\left( x_0/x\right)}
{\ln\left(\ln\left(Q^2/\Lambda^2\right)/\ln\left(Q_0^2/\Lambda^2\right)\right)}
\right]^{1/2}
\,\mbox{.}
\label{eq:das2}
\end{equation}
The parameters $x_0$ and $Q_0^2$ define the formal boundaries of the asymptotic
region, ${\cal{N}}_\Sigma$ and ${\cal{N}}_g$ are normalization constants,
and $n_f$ is the number of active flavours. 
Using the asymptotic form Eq.~(\ref{eq:das}) in the definition of the 
effective exponents
Eq.~(\ref{eq:def}) then gives us a perturbative prediction for the small-$x$ 
exponents $\alpha_{\Sigma}$ and $\alpha_g$: at large $\sigma$ but fixed $\rho$ one has
\begin{equation}
\alpha_{\Sigma}(x,Q^2)
\to -\frac{\gamma}{\rho}+\frac{3}{4\sigma\rho}\, ,\qquad
\alpha_{g}(x,Q^2)
\to -\frac{\gamma}{\rho}+ \frac{1}{4\sigma\rho}\, .
\label{eq:dasvalues}
\end{equation}
Note that both $\alpha_{\Sigma}(x,Q^2)$ and $\alpha_g(x,Q^2)$ converge 
asymptotically to the same value $-\gamma/\rho$, as expected since the 
QCD evolution of the gluon distribution seeds the evolution of the quark 
singlet distribution.
The DAS results Eq.~(\ref{eq:dasvalues}),
which are a generic prediction of perturbative QCD, 
are displayed in
Fig.~\ref{fig:evolsinglet}, where we have used $x_0=0.1$, $Q_0^2=1$ GeV$^2$, $n_f=5$ 
and $\Lambda^{(n_f=5)}=0.220$ GeV.
The agreement between the expectation from DAS and results from the global 
fits is excellent at 
$Q^2\gtrsim 10$ GeV$^2$ for both the quark singlet and the gluon.

At large $x$, the $Q^2$ dependence of the effective exponents
can also be determined from general perturbative QCD considerations, 
following directly from 
the universality of the cusp quark anomalous dimension in the 
$\overline{{\rm MS}}$ scheme~\cite{Korchemsky:1988si,Albino:2000cp}. 
Specifically, it can be shown, either by analysing Wilson 
lines~\cite{Korchemsky:1988si}, or by using standard 
results for the exponentiation of soft logarithms in the quark-initiated 
bare cross sections~\cite{Albino:2000cp}, that the quark anomalous dimension at 
large $N$ takes the universal form
\begin{equation}
\gamma_q(N,\alpha_s(q^2))\sim 
- c(\alpha_s(q^2))\ln N + d(\alpha_s(q^2))+O(1/N) \, ,
\end{equation}
where 
$c(\alpha_s(q^2))$ and $d(\alpha_s(q^2))$  can be computed perturbatively: 
for example at NLO 
\begin{equation}
c(\alpha_s(q^2)) = \frac{\alpha_s(q^2)}{2\pi}c_1 
+ \left(\frac{\alpha_s(q^2)}{2\pi}\right)^2 c_2 + O(\alpha_s^3) \, ,
\label{eq:Q}
\end{equation}
with coefficients~\cite{Kodaira:1981nh}
\begin{equation}
c_1 = \frac{8}{3}\, ,\qquad
c_2 = 4\left(\frac{67}{9} - 2\zeta_2\right) - \frac{40}{27} n_f \, .
\end{equation}
It follows~\cite{Albino:2000cp} that, if 
$xf_{q}(x,Q_0^2)\sim (1-x)^{b(Q_0^2)}$ as $x\to 1$ at a scale $Q_0^2$,
with $q$ either the quark
singlet, $\Sigma$, or one of the quark valence distributions, $u_V$ or $d_V$, 
then this asymptotic behaviour persists at higher scales $Q^2$ with 
\begin{equation}
b(Q^2) = b(Q_0^2)+\int_{Q_0^2}^{Q^2}\frac{dq^2}{q^2}c(\alpha_s(q^2))\, .
\label{eq:bscale}
\end{equation}
Given our definition Eq.~(\ref{eq:def}) and the asymptotic behaviour 
Eq.~(\ref{eq:defasybeta}) at large $x$, as $x\to 1$ one has
\begin{equation}
\beta_{f_i}(x,Q^2) 
= 
\beta_{f_i}(x,Q_0^2)+\int_{Q_0^2}^{Q^2}\frac{dq^2}{q^2}c(\alpha_s(q^2))\, .
\label{eq:betascale}
\end{equation}
The behaviour predicted by Eq.~(\ref{eq:betascale}) is displayed for $u_V$ and 
$d_V$ in Fig.~\ref{fig:evolvalence}, and for $\Sigma$
in Fig.~\ref{fig:evolsinglet}. Note that Eq.~(\ref{eq:betascale}) 
only determines the shape of the curve, not its overall normalization;
for definiteness we fix the value of $\beta(x,Q_0^2)$ in 
Eq.~(\ref{eq:betascale}) to match the central 
values obtained from {\tt CT14} at $Q^2=10^6$ GeV$^2$. 
The agreement between Eq.~(\ref{eq:betascale}) and the $Q^2$ dependence of the 
large-$x$ effective exponents derived from the 
PDF fit is excellent. A slight deterioration only appears at small values of 
$Q^2$ due to missing higher order corrections. Similar conclusions can be derived for
other 
PDF sets when the value of $\beta_{f_i}(x,Q_0^2)$ in Eq.~(\ref{eq:betascale})
is assigned consistently.

%% file: sec-comparison.tex
\section{Comparison with nonperturbative predictions}
\label{sec:comparisons}

We now discuss how our findings compare with the 
expectations from Regge theory and the Brodsky-Farrar quark counting rules.
In Tabs.~\ref{tab:alphavalues}-\ref{tab:betavalues} 
we show the values of the effective exponents
for the {\tt NNPDF3.0}, {\tt CT14}, {\tt MMHT14}, {\tt ABM12} and {\tt CJ15} 
PDF sets, computed at $x_a=10^{-4}$ and $x_b=0.9$ ($x_b=0.5$ for $S$)
at $Q^2=2$ GeV$^2$ and $Q^2=10$ GeV$^2$.
We also include the values predicted by Regge theory and the Brodsky-Farrar 
quark counting rules.

\begin{table}[!t]
\centering
\small
\begin{tabular}{cccccccc}
\toprule
\multirow{2}{*}{$f_i$}                    &  
$Q^2$                                     & 
\multicolumn{5}{c}{$\alpha_{f_i}(x_a,Q^2)$} & 
\multirow{2}{*}{$a_{f_i}$}\\[0.1cm]
               & 
[GeV$^2$]      & 
{\tt NNPDF3.0} & 
{\tt CT14}     & 
{\tt MMHT14}   & 
{\tt ABM12}    & 
{\tt CJ15}\\
\midrule
\multirow{2}{*}{$u_V$} &
$2.0$ & 
$+0.48\pm 0.11$ & 
$+0.72\pm 0.12$ & 
$+0.65\pm 0.06$ & 
$+0.76\pm 0.07$ & 
$+0.61\pm 0.01$ & 
$+0.5$\\
& $10.0$ & 
$+0.46\pm 0.09$ & 
$+0.66\pm 0.09$ & 
$+0.61\pm 0.04$ & 
$+0.70\pm 0.04$ & 
$+0.60\pm 0.01$ & 
($0.63$)\\
\midrule
\multirow{2}{*}{$d_V$} &
$2.0$ & 
$+0.41\pm 0.11$ & 
$+0.73\pm 0.12$ & 
$+0.79\pm 0.06$ & 
$+1.39\pm 0.10$ & 
$+1.11\pm 0.03$ & 
$+0.5$\\
& $10.0$ & 
$+0.41\pm 0.11$ & 
$+0.66\pm 0.07$ & 
$+0.70\pm 0.04$ & 
$+0.91\pm 0.08$ & 
$+0.95\pm 0.05$ &
($0.63$) \\
\midrule
\multirow{2}{*}{$S$} &
$2.0$ & 
$-0.14\pm 0.06$ & 
$-0.15\pm 0.05$ & 
$-0.09\pm 0.04$ & 
$-0.16\pm 0.02$ &
$-0.18\pm 0.03$ & 
$-0.08$\\
& 
$10.0$ & 
$-0.18\pm 0.04$ & 
$-0.20\pm 0.05$ & 
$-0.15\pm 0.04$ & 
$-0.19\pm 0.01$ & 
$-0.14\pm 0.02$ & 
($-0.2$) \\
\midrule
\multirow{2}{*}{$g$} &
$2.0$ & 
$-0.16\pm 0.63$ & 
$+0.06\pm 0.31$ & 
$-0.79\pm 0.43$ & 
$+0.18\pm 0.10$ & 
$+0.08\pm 0.03$ & 
$-0.08$\\
& 
$10.0$ & 
$-0.20\pm 0.46$ & 
$-0.15\pm 0.15$ & 
$-0.29\pm 0.09$ & 
$-0.15\pm 0.01$ & 
$-0.14\pm 0.01$ & 
($-0.2$)\\
\bottomrule
\end{tabular}
\normalsize
\caption{\small The values of the small-$x$ effective exponent 
$\alpha_{f_i}(x_a,Q^2)$ computed at $Q^2=2$ GeV$^2$ and $Q^2=10$ GeV$^2$ at 
$x_a=10^{-4}$, compared to the values of $a_{f_i}$ predicted by Regge theory
(and resummation of double logarithms).
For the quark sea $S$ and the gluon $g$ we indicate the prediction 
of the soft Pomeron (and the NLLx perturbative result).}
\label{tab:alphavalues}
\end{table}

\begin{table}[!t]
\centering
\small
\begin{tabular}{cccccccc}
\toprule
\multirow{2}{*}{$f_i$} &  
$Q^2$ & 
\multicolumn{5}{c}{$\beta_{f_i}(x_b,Q^2)$} & 
\multirow{2}{*}{$b_{f_i}$}\\[0.1cm]
& [GeV$^2$] & 
{\tt NNPDF3.0} &
{\tt CT14}     & 
{\tt MMHT14}   & 
{\tt ABM12}    & 
{\tt CJ15}\\
\midrule
\multirow{2}{*}{$u_V$} &
$2.0$ & 
$+2.94\pm 0.52$ & 
$+3.11\pm 0.28$ & 
$+3.37\pm 0.07$ & 
$+3.38\pm 0.06$ & 
$+3.50\pm 0.01$ & 
\multirow{2}{*}{$\sim 3$}\\
& 
$10.0$ & 
$+3.30\pm 0.69$ & 
$+3.38\pm 0.29$ & 
$+3.62\pm 0.07$ & 
$+3.61\pm 0.05$ & 
$+3.78\pm 0.01$ & \\
\midrule
\multirow{2}{*}{$d_V$} &
$2.0$ & 
$+3.03\pm 1.96$ & 
$+3.27\pm 0.37$ & 
$+2.05\pm 0.59$ & 
$+4.72\pm 0.43$ & 
$+3.42\pm 0.06$ & 
\multirow{2}{*}{$\sim 3$}\\
& 
$10.0$ & 
$+3.23\pm 1.88$ & 
$+3.52\pm 0.36$ & 
$+2.29\pm 0.59$ & 
$+4.92\pm 0.42$ & 
$+3.68\pm 0.05$ &\\
\midrule
\multirow{2}{*}{$S$} &
$2.0$ & 
$+6.86 \pm 7.25$ & 
$+6.41 \pm 1.22$ & 
$+8.19 \pm 0.68$ & 
$+8.16 \pm 0.38$ & 
$+7.73 \pm 0.18$ & 
\multirow{2}{*}{$\sim 7$}\\
& 
$10.0$ & 
$+6.76 \pm 6.71$ & 
$+6.91 \pm 1.14$ & 
$+6.83 \pm 0.88$ & 
$+8.51 \pm 0.38$ & 
$+8.15 \pm 0.18$ & \\
\midrule
\multirow{2}{*}{$g$} &
$2.0$ & 
$+2.95\pm 1.25$ & 
$+5.08\pm 2.18$ & 
$+1.65\pm 0.23$ & 
$+4.18\pm 0.06$ & 
$+6.11\pm 0.33$ & 
\multirow{2}{*}{$\sim 5$}\\
& 
$10.0$ & 
$+3.25\pm 0.98$ & 
$+5.13\pm 0.51$ & 
$+2.24\pm 0.23$ & 
$+4.44\pm 0.06$ & 
$+4.91\pm 0.33$ & \\
\bottomrule
\end{tabular}
\caption{\small Same as Tab.~\ref{tab:alphavalues}
for the large-$x$ effective exponent
$\beta_{f_i}(x_b,Q^2)$ at $x_b=0.9$ (for $u_V$, $d_V$ and $g$) and 
$x_b=0.5$ (for $S$). The values of the exponent $b_{f_i}$
predicted by Brodsky-Farrar quark counting rules are also shown.}
\label{tab:betavalues}
\end{table}

At small $x$, Regge theory predicts $xf_i\sim x^{a_{f_i}}$ with $a_{f_i}$ a 
$Q^2$-independent exponent, related to the intercept of the corresponding 
Regge trajectory. 
For valence quark distributions, a value of 
$a_{u_V}=a_{d_V}\simeq +0.5$ is derived from the non-singlet Regge trajectory 
intercept $1-\alpha_R(0)$. Perturbative calculations which resum the double logarithms of $x$ give a similar value $a_{u_V}=a_{d_V}\simeq +0.63$ 
\cite{Ermolaev:2000sg, Ermolaev:2009cq}. 
For the gluon distribution, a value of $a_g$ close 
to the singlet Pomeron trajectory $1-\alpha_P(0)$ is expected; the 
conventional Regge exchange is that of the soft 
Pomeron~\cite{Abarbanel:1969eh} (for a formulation of the parton picture 
without recourse to perturbation theory see also Ref.\cite{Landshoff:1970ff}), 
leading to 
$a_g\simeq -0.08$. Attempts to compute the Pomeron intercept perturbatively  
by solution of the fixed coupling LLx BFKL 
equation~\cite{Lipatov:1976zz,Fadin:1975cb,Kuraev:1976ge,Kuraev:1977fs} 
give $a_g\simeq -0.5$. However 
this result is destabilised by NLLx corrections~\cite{Fadin:1998py}. 
When running 
coupling effects are taken into account, the perturbative expansion is 
stabilised~\cite{Altarelli:2001ji,Altarelli:2003hk,Ciafaloni:2003rd,Altarelli:2005ni,Ciafaloni:2007gf}, and the 
NLLx perturbative prediction becomes  $a_g\simeq -0.2$. 
For the total sea distribution, the value of $a_S$
should be similar for large enough $Q^2$ to $a_g$,
due to the dominance of the
process $g\to q\bar{q}$ in the evolution 
of sea quarks.

In comparing these expectations with the results from PDF fits, 
we need to choose a scale. 
Regge predictions are expected to hold only at low scales. 
For $\alpha_{u_V}(x,Q^2)$ and 
$\alpha_{d_V}(x,Q^2)$ this is not too much of a problem, since the scale 
dependence of non-singlet distributions is quite weak (see Fig.~\ref{fig:evolvalence}). 
The values extracted from {\tt NNPDF3.0} are 
accordingly in good agreement with Regge expectations; 
those from the other global PDF fits are 
generally a little high (see Tab.~\ref{tab:alphavalues}).  
On the other hand, for 
$\alpha_S(x,Q^2)\simeq \alpha_\Sigma(x,Q^2)$ and $\alpha_g(x,Q^2)$, 
the scale dependence is 
rather strong (see Fig.~\ref{fig:evolsinglet}), 
due to the double scaling behaviour. Making 
the comparison at low scales, we see reasonable agreement 
for the sea quarks with 
the Pomeron prediction, and also with the NLLx perturbative prediction. 
Uncertainties for 
the gluon intercept are inevitably large, so here the agreement 
is only qualitative. Note 
that for {\tt ABM12} and
{\tt CJ15} the uncertainties are 
often substantially 
underestimated due to parametrisation constraints in the 
extrapolation region.

\begin{figure}[!t]
\begin{center}
\begin{fmffile}{CountingRules}
{
\ \ \ \
\savebox\feynbox{
\begin{fmfgraph*}(30,20)
\fmfstraight
\fmfbottom{il,ii,ir}
\fmftop{itl1,itl2,itl3}
\fmfv{lab=(a)}{itl1}
\fmf{plain}{il,ii}
\fmf{fermion}{ii,ir}
\fmf{photon,label=$\gamma^*$}{itl1,ii}
\fmfi{plain}{vpath(__il,__ii) shifted (thick*(0,-4))}
\fmfi{plain_arrow}{vpath(__ii,__ir) shifted (thick*(0,-4))}
\fmfi{plain}{vpath(__il,__ii) shifted (thick*(0,-8))}
\fmfi{plain_arrow}{vpath(__ii,__ir) shifted (thick*(0,-8))}
\end{fmfgraph*}
}
\usebox{\feynbox}\postfeynlabel{-1.6}{.2}{\feynbox}{$n_s=2$}
\ \ \ \
\savebox\feynbox{
\begin{fmfgraph*}(30,20)
\fmfstraight
\fmfbottom{il,ii,ir}
\fmftop{itl1,itl,itl3}
\fmfv{lab=(b)}{itl1}
\fmfright{itr1,itr2,itr3,itr4,itr5}
\fmf{plain}{il,ii}
\fmf{fermion}{ii,ir}
\fmf{photon,label=$\gamma^*$}{itl,itr3}
\fmf{gluon}{ii,itr3}
\fmfi{plain}{vpath(__il,__ii) shifted (thick*(0,-4))}
\fmfi{plain_arrow}{vpath(__ii,__ir) shifted (thick*(0,-4))}
\fmfi{plain}{vpath(__il,__ii) shifted (thick*(0,-8))}
\fmfi{plain_arrow}{vpath(__ii,__ir) shifted (thick*(0,-8))}
\fmfblob{.1w}{itr3}
\fmfv{lab=BGF,lab.dist=0.08w}{itr3}
\end{fmfgraph*}
}
\usebox{\feynbox}\postfeynlabel{-.8}{.4}{\feynbox}{$n_s=3$}
\ \ \ \
\savebox\feynbox{
\begin{fmfgraph*}(30,20)
\fmfstraight
\fmfbottom{il,ii,ir}
\fmftop{itl1,itl,itl3}
\fmfv{lab=(c)}{itl1}
\fmfright{itr1,itr2,itr3,itr4,itr5,itr6}
\fmf{plain}{il,ii}
\fmf{fermion}{ii,ir}
\fmf{gluon,tension=0.9}{ii,v}
\fmf{fermion}{v,itr3}
\fmf{photon,tension=0.3,label=$\gamma^*$}{itl,vp}
\fmf{fermion}{vp,v}
\fmf{fermion}{itr6,vp}
\fmfi{plain}{vpath(__il,__ii) shifted (thick*(0,-4))}
\fmfi{plain_arrow}{vpath(__ii,__ir) shifted (thick*(0,-4))}
\fmfi{plain}{vpath(__il,__ii) shifted (thick*(0,-8))}
\fmfi{plain_arrow}{vpath(__ii,__ir) shifted (thick*(0,-8))}
\end{fmfgraph*}
}
\usebox{\feynbox}\postfeynlabel{-.4}{.8}{\feynbox}{$n_s=4$}
}
\end{fmffile}
\end{center}
\vspace{-0.4cm}
\caption{\small The number of {\it spectator} partons in a proton consisting of 
three quarks, whether a valence quark (a), a gluon (b) or a sea quark (c) 
is struck by a virtual photon $\gamma^*$ in deep-inelastic scattering.}
\label{fig:CRgraphs}
\end{figure}
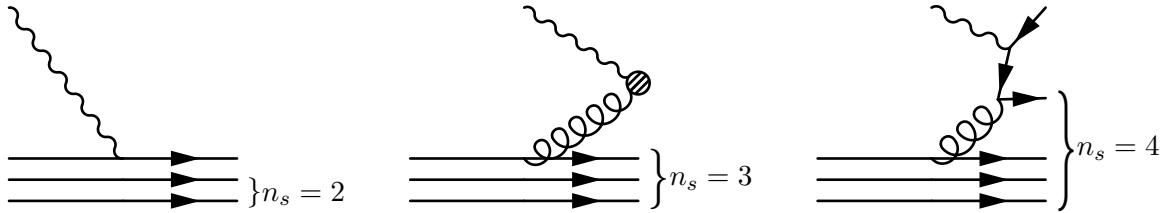

At large $x$, the Brodsky-Farrar quark counting rules predict that 
$xf_i\sim (1-x)^{2n_s-1}$, where $n_s$ is the minimum number of {\it spectator}
partons. These are defined to be the partons that are not struck in the hard 
scattering process,
since it is assumed that in the limit $x\to 1$, there can be no momentum left 
for any of the partons other than the struck parton. In a proton 
made of three quarks, one has: for a valence quark, $n_s=2$ and thus 
$b_{u_V}=b_{d_V}=3$; 
for a gluon, $n_s=3$ and $b_g=5$; for a sea quark, $n_s=4$ and  
$b_S=7$; see Fig.~\ref{fig:CRgraphs}. Note that the values of the exponents 
predicted by Brodsky-Farrar quark counting rules are different if the 
polarization of the quark with respect to the polarization of the parent 
hadron is retained~\cite{Brodsky:1994kg}. This also affects the 
difference between up and down distributions. 
A detailed comparison between PDFs and 
quark counting rules in the polarized case was presented in 
Ref.~\cite{Nocera:2014uea}.
Again it is unclear from the quark model argument 
at which scale these predictions are supposed to apply, 
but again we are fortunate that 
the scale dependence of large-$x$ PDFs is reasonably
moderate (see Fig.~\ref{fig:evolvalence} and 
Fig.~\ref{fig:evolsinglet}), and it is reasonable to make the comparison at 
a low scale~\cite{Deur:2016cxb}.

The predictions $b_{u_V}(x,Q^2)$ and 
$b_{d_V}(x,Q^2)$ for the valence distributions are then in broad agreement with 
the effective exponents determined from most of the global PDF fits, 
though some deviations from Brodsky-Farrar quark counting rule 
expectations are observed for the {\tt MMHT14} down valence quarks: 
this seems to be a 
result of the oscillation noted already in Fig.~\ref{fig:expeffval}. 
For the quark sea and the gluon, the success is again rather mixed, 
and only {\tt CT14} seems to 
provide results which agree with the prediction;  
for {\tt NNPDF3.0} the uncertainties on the quark sea are too large for 
the extraction 
to be meaningful, while the result for the gluon is a little low; 
for {\tt MMHT14} the result for the gluon is far too low, 
with a substantially underestimated uncertainty.

In addition to the Brodsky-Farrar quark counting rules, 
the behaviour of PDFs at 
large $x$ has been predicted by several nonperturbative models of 
nucleon structure (see {\it e.g.}~\cite{Melnitchouk:1995fc,Holt:2010vj} 
and references therein). In many cases, these provide expectations for 
the ratio of $u$ to $d$ valence distributions in the proton, $d_V/u_V$, 
and of neutron to proton structure functions, $F_2^n/F_2^p$.
These ratios are particularly interesting because while all PDFs vanish at 
$x=1$, their ratio does not necessarily do so, and thus
it is a useful discriminator among models of nucleon structure.

In the parametrisation Eq.~(\ref{eq:parametrisation}), 
$d_V/u_V\sim (1-x)^{b_{d_V}-b_{u_V}}$ as $x\to 1$, so if $b_{u_V}=b_{d_V}$, as predicted 
by the counting rules, then $d_V/u_V\to k$, with $k$ some constant. 
Indeed it is the constant $k$ that many of the models try to predict. 
Moreover, as noted above, 
both {\tt CT14} and {\tt CJ15} assume  $b_{u_V}=b_{d_V}$ in their fits. 
However while one may expect $b_{u_V}\simeq b_{d_V}$ because of isospin symmetry,
it is also reasonable to expect that exact equality will be broken by isospin 
breaking or electromagnetic effects. The sign of these 
effects is crucial: if $b_{u_V}>b_{d_V}$ then $d_V/u_V$ will become infinite 
as $x\to 1$, while 
if $b_{u_V}<b_{d_V}$, as $x\to 1$ $d_V/u_V\to 0$. These two possibilities result 
in naive limits on the ratio $F_2^n/F_2^p$: if the sea quarks can be ignored 
at large $x$, then $d_V\gg u_V$, $F_2^n/F_2^p\to 4$, 
while for $d_V\ll u_V$ $F_2^n/F_2^p\to 1/4$, giving for $x\to 1$ the 
Nachtmann limits~\cite{Nachtmann:1972pc}
\begin{equation}
\frac{1}{4}\leq\frac{F_2^n}{F_2^p}\leq 4\, .
\label{eq:nachtmann}
\end{equation}

\begin{figure}[!t]
\centering
\includegraphics[width=0.46\textwidth,angle=270]{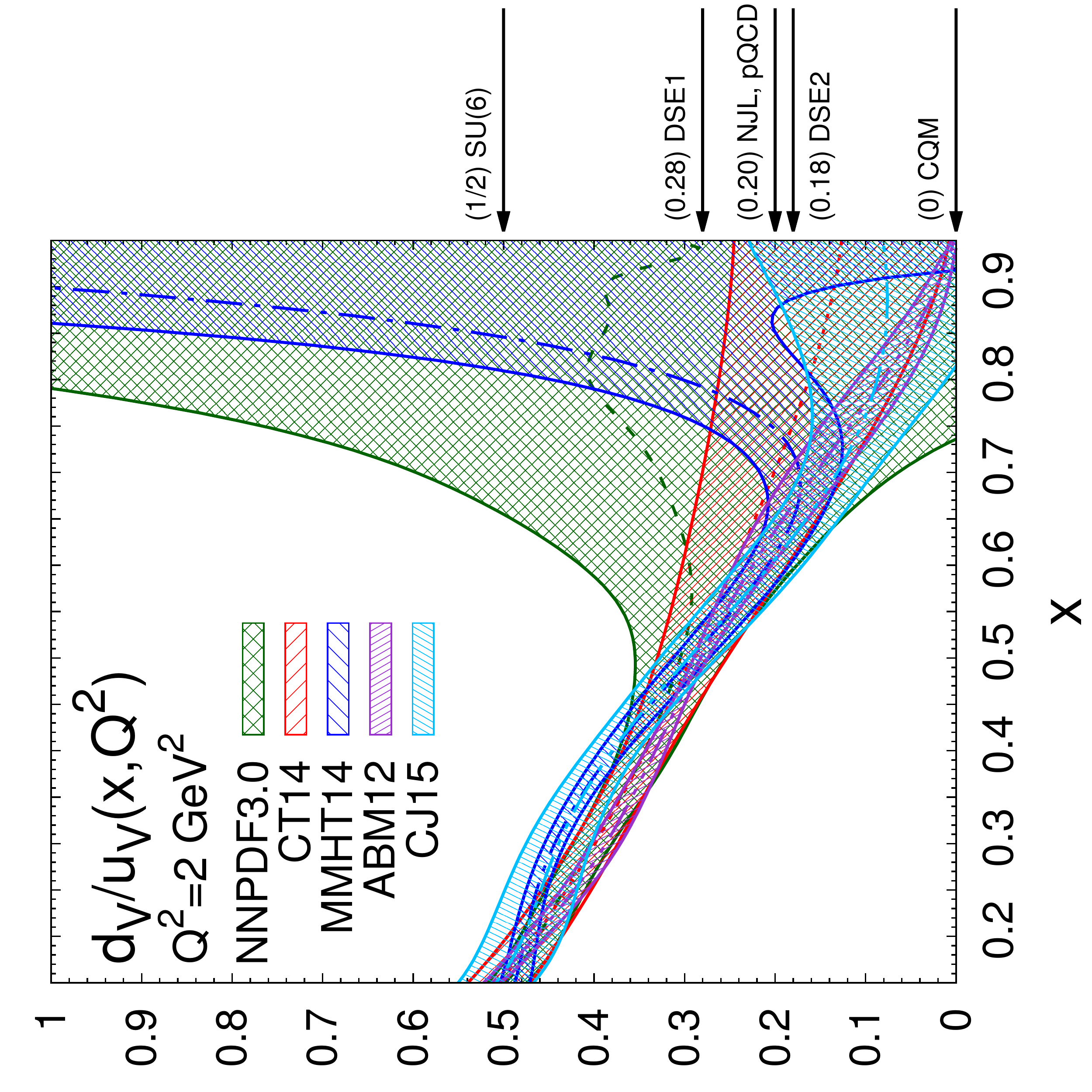}
\includegraphics[width=0.46\textwidth,angle=270]{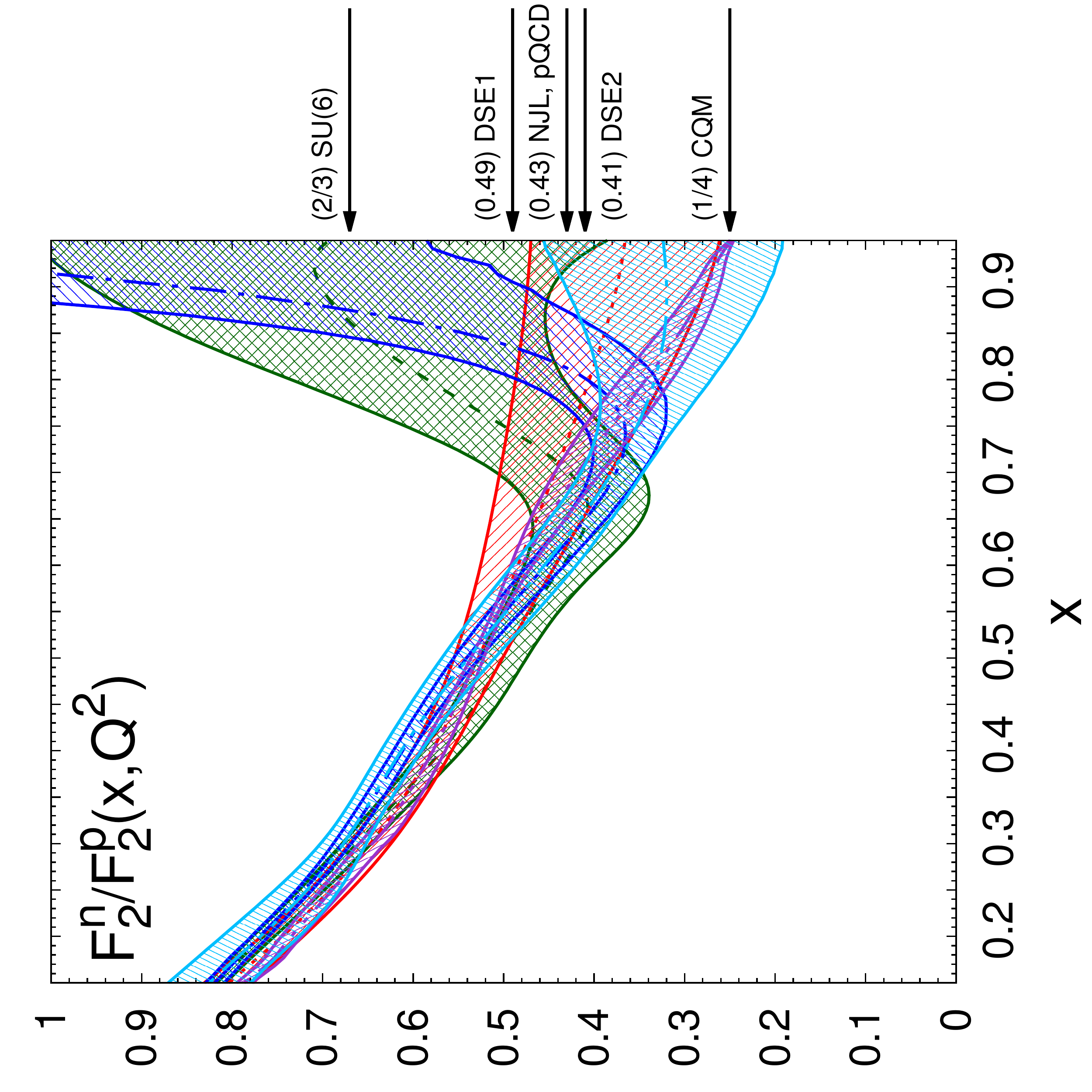}
\caption{\small The ratios $d_V/u_V$ (left) and $F_2^n/F_2^p$ (right) at 
$Q^2=2$ GeV$^2$ among various PDF sets, compared with the predictions of 
different nonperturbative models of nucleon structure.}
\label{fig:ratioplots}
\end{figure}

To address these issues empirically, in Fig.~\ref{fig:ratioplots} we compare 
the ratios $d_V/u_V(x,Q^2)$ 
and $F_2^n(x,Q^2)/F_2^p(x,Q^2)$ at $Q^2=2$ GeV$^2$ as predicted by the various 
PDF sets.
The neutron and proton structure functions $F_2^n(x,Q^2)$ and $F_2^p(x,Q^2)$ 
have been computed 
at NNLO accuracy with {\tt APFEL}~\cite{Bertone:2013vaa} using the FONLL-C
general-mass scheme~\cite{Forte:2010ta}.
The arrows on the right hand side of each panel indicate
the expectations from a representative set of nonperturbative models
of nucleon structure:
${\rm\tt SU(6)}$~\cite{Close:1979bt} describes constituent quarks 
in the nucleon by ${\rm SU(6)}$ wave functions; 
{\tt CQM}~\cite{Close:1973xw,Carlitz:1975bg} is the relativistic 
Constituent Quark Model in which a ${\rm SU(6)}$ symmetry breaking is assumed
via a color hyperfine interaction between quarks; 
{\tt NJL}~\cite{Cloet:2005pp} is a modified Nambu--Jona-Lasinio model in
which confinement is simulated by eliminating unphysical thresholds for 
nucleon decay; {\tt pQCD}~\cite{Farrar:1975yb} stands for a 
coloured quark and vector gluon model supplemented with leading order
perturbative QCD; {\tt DSE1} and {\tt DSE2}~\cite{Roberts:2013mja} are two 
scenarios based on Dyson-Schwinger equations.

From Fig.~\ref{fig:ratioplots}, we see that in the region in which the
valence quarks are 
constrained by experimental data, {\it i.e.} $x\lesssim 0.5$, 
the predictions for both ratios 
from all the PDF sets are in reasonable agreement with each other within uncertainties, 
as might be expected. 
For $x\gtrsim 0.5$, the mutual consistency of PDF sets deteriorates 
rapidly, and a wide range of different behaviours is observed. 
This is a consequence of the 
reduced experimental information in this region: different PDF collaborations 
extrapolate 
to large $x$ using different assumptions. For those sets with very weak 
assumptions on the 
PDF behaviour at large $x$, namely {\tt NNPDF3.0} and {\tt MMHT14}, 
the uncertainties on 
the ratios expand rapidly, and at very large $x$ there is no predictive power at all. 
For the two sets which assume that $d_V/u_V\to k$ at large $x$, namely 
{\tt CT14} and {\tt CJ15},
uncertainties are inevitably much reduced and a value of $k$ is predicted. 
{\tt ABM12} is different again, in that they find as a result of their fit 
that $b_{d_V}>b_{u_V}$ 
at more than two standard deviations (see Tab.~\ref{tab:betavalues}), 
so that $d_V/u_V\to 0$ as $x\to 1$, and an unrealistically small uncertainty 
band in a region where there are actually no data. 

It follows that all the various model predictions displayed 
in Fig.~\ref{fig:ratioplots} 
are compatible with the {\tt NNPDF3.0} and {\tt MMHT14} predictions, 
while {\tt ABM12} confirms 
the Chiral Quark Model but appears to rule out all the others.
The {\tt CT14} and {\tt CJ15} sets favour values of $k$ in the region 
$0\lesssim k \lesssim 0.25$,
thus disfavouring the $SU(6)$ prediction but unable to discriminate between 
the others. 
The preference for smaller values of $k$ results in effect from a linear 
extrapolation of the 
downwards trend in the data region $x\lesssim 0.5$. Not all the predictions 
respect the Nachtmann bound Eq.~(\ref{eq:nachtmann}).

%% file: sec-conclusion.tex
\section{Conclusions and outlook}
\label{sec:conclusion}

In summary, in this work we have introduced a novel
methodology  to determine quantitatively
the effective asymptotic behaviour of parton distributions, valid for any 
value of $x$ and 
$Q^2$. For the first time, we have unambiguously identified
the ranges in $x$ and $Q^2$ where the asymptotic regime sets in,
allowing us to 
compare in detail perturbative and nonperturbative 
QCD predictions at large and small $x$ with the results of modern 
global PDF fits.

Concerning the small-$x$ region, we have found broad agreement between 
the results from PDF fits and the predictions from Regge theory for the 
behaviour of the 
valence quark distributions. For the singlet and gluon distributions, 
the agreement with Regge predictions  is still only qualitative, 
due in part to the substantial scale dependence, as well as the limited
experimental information available in that region.
On the other hand, the perturbative QCD Double Asymptotic Scaling
predictions are in excellent agreement with the results of PDFs fits
over a wide range of $Q^2$. 

Concerning the large-$x$ region, we have found that the 
predictions of the Brodsky-Farrar counting rules for the behaviour of the 
valence quark distributions are in broad agreement with 
the global fit results, within PDF uncertainties. For the sea and gluon 
distributions uncertainties are much larger, and the agreement is only 
qualitative. The scale dependence of the effective exponents
based on global PDF fits is in excellent agreement with the perturbative QCD
expectation from the cusp anomalous dimension in a wide range of $Q^2$.
We have also compared the ratios $d_V(x,Q^2)/u_V(x,Q^2)$ and 
$F_2^n(x,Q^2)/F_2^p(x,Q^2)$ among PDF fits and with nonperturbative models
of nucleon structure, but found that the interpretation of this comparison 
depends significantly on the assumptions built into the PDF 
parametrisation, 
to the extent that it is impossible at present to draw any firm conclusions.

We therefore conclude that, while the ancient wisdom of Regge theory 
and the Brodsky-Farrar counting rules seems to have some degree of truth,
particularly in the valence quark 
sector, they are no substitute for the precise empirical PDF determinations 
provided by global analysis, and when used as constraints may lead to 
unrealistically accurate predictions in kinematic regions where there is no experimental data. 
Global PDF fits will always be hampered to some extent 
by the lack of data to constrain PDFs in extrapolation regions,
and new  measurements from the LHC and other facilities, 
such as JLab, are required to shed more light on the asymptotic behaviour 
of parton
distributions at small and large $x$. The methodology presented in this work
should find applications in future comparisons between different global PDF 
fits, and between PDF fits and nonperturbative models of nucleon structure.

%% file: sec-appendix.tex
\section*{Appendix: numerical determination of the effective exponents}
\label{sec:appendix}

The accurate evaluation of the effective exponents $\alpha_{f_i}(x,Q^2)$ and 
$\beta_{f_i}(x,Q^2)$ through Eq.~(\ref{eq:def}) is pivotal in our study.
An analytic evaluation of Eq.~(\ref{eq:def}) starting from the explicit PDF
parametrisation in Eq.~(\ref{eq:parametrisation}), though straightforward, has
two main limitations. First, Eq.~(\ref{eq:parametrisation}) holds only at the 
initial parametrisation scale $Q_0^2$; the form of 
Eq.~(\ref{eq:parametrisation}) is rapidly washed out by DGLAP
evolution, hence it cannot be used for the analytic computation of 
the effective exponents at $Q^2 > Q_0^2$. Second, even at $Q^2=Q_0^2$, only the 
best-fit parameters for the central PDF are provided, and moreover for some 
PDF fits not even a simple analytical parametrisation is used. 

To overcome these difficulties, in this work we evaluate Eq.~(\ref{eq:def}) 
numerically. To this purpose, PDFs in a suitable numerical format  
and an algorithm for the numerical computation of the logarithmic derivative 
of the PDF in Eq.~(\ref{eq:def}) are necessary. The first requirement is 
fulfilled by {\tt LHAPDF6}~\cite{Buckley:2014ana}, while the 
second is more delicate.
The standard methods used to evaluate numerical derivatives,
such as those based on a finite difference approximation or on a 
polynomial approximation of the function to be derived,
see {\it e.g.} Sec.~5.7-5.9 in Ref.~\cite{Press:1992zz},
are found to lead to unstable results.
The reason is that PDFs available through {\tt LHAPDF6} are
tabulated on a grid in $(x,Q^2)$; the values of the PDFs off a grid node 
are then obtained by a cubic spline interpolation. This interpolation induces
small fluctuations of the PDFs with respect to their {\it true} value,
in particular in the small and large-$x$ regions, where the grid tabulations 
are less dense. Such fluctuations are enhanced when the numerical derivative 
of the PDF is computed, especially if the value of the PDF is very small, thus spoiling the 
evaluation of Eq.~(\ref{eq:def}).

We overcome this problem and perform the numerical derivative in 
Eq.~(\ref{eq:def}) by means of a Savitzky-Golay smoothing 
filter~\cite{doi:10.1021/ac60214a047}. The idea is the following.
Assuming that a function $g(x)$ 
is tabulated at $n+1$ equally spaced intervals, $g_i\equiv g(x_i)$,
with $x_i=x_0+i\Delta$ for some constant sample spacing $\Delta=(x_n-x_0)/n$ 
and $i=-n/2,\dots,-2,-1,0,1,2,\dots,n/2$,  
the filter performs a least-squares fit with a polynomial of some degree $m$ 
at each point, using an additional number $n_L$ of points to the left and 
some number $n_R$ of points to the right of each desired $x$ value.
The estimated derivative is then the derivative of the resulting
fitted polynomial.
The values of the parameters $x_0$, $x_n$, $n$, $n_L$, $n_R$ and $m$ 
are optimized for each flavour and PDF set, so that residual numerical 
instabilities are minimised.

\begin{figure}[!t]
\centering
\includegraphics[width=0.495\textwidth,angle=270]{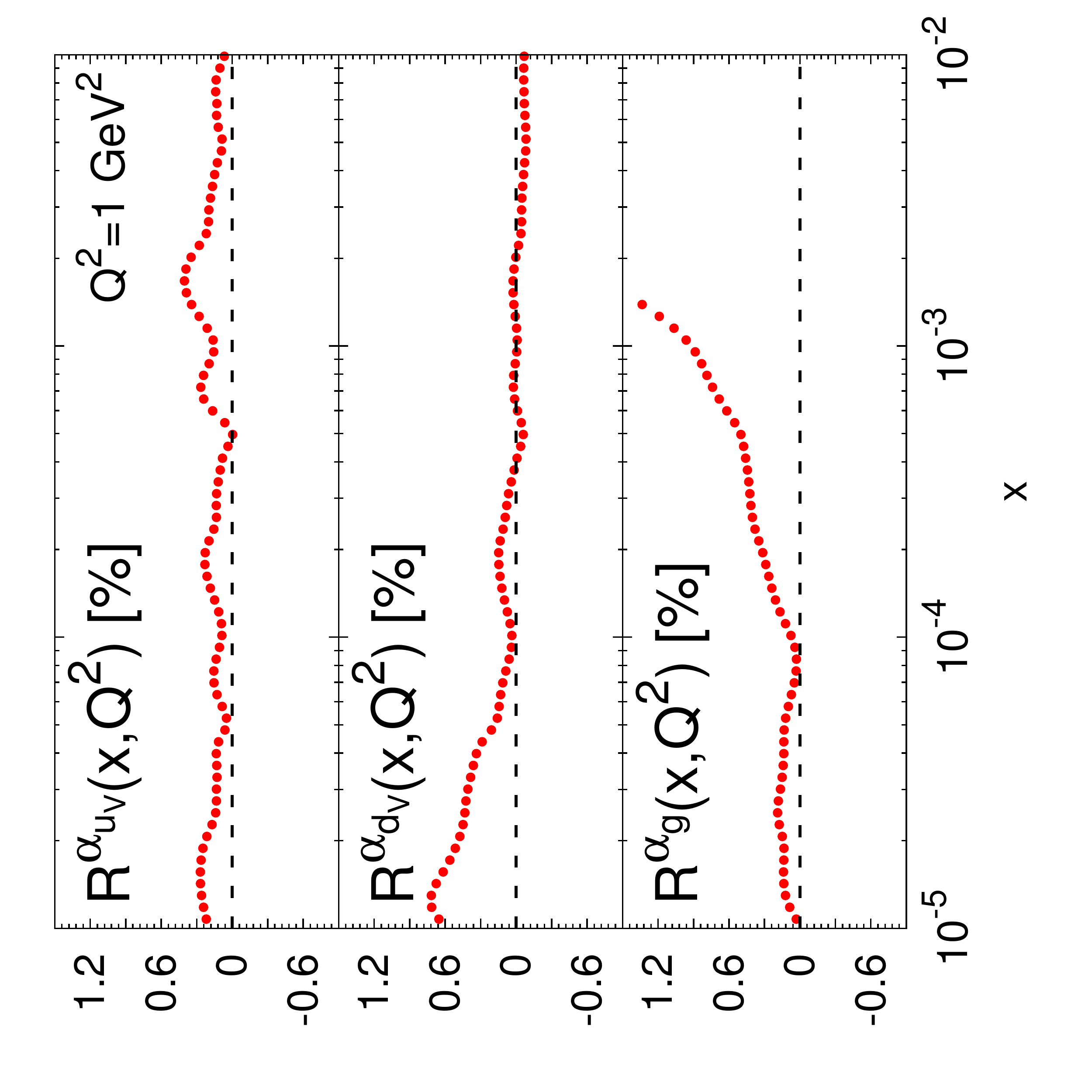}
\includegraphics[width=0.495\textwidth,angle=270]{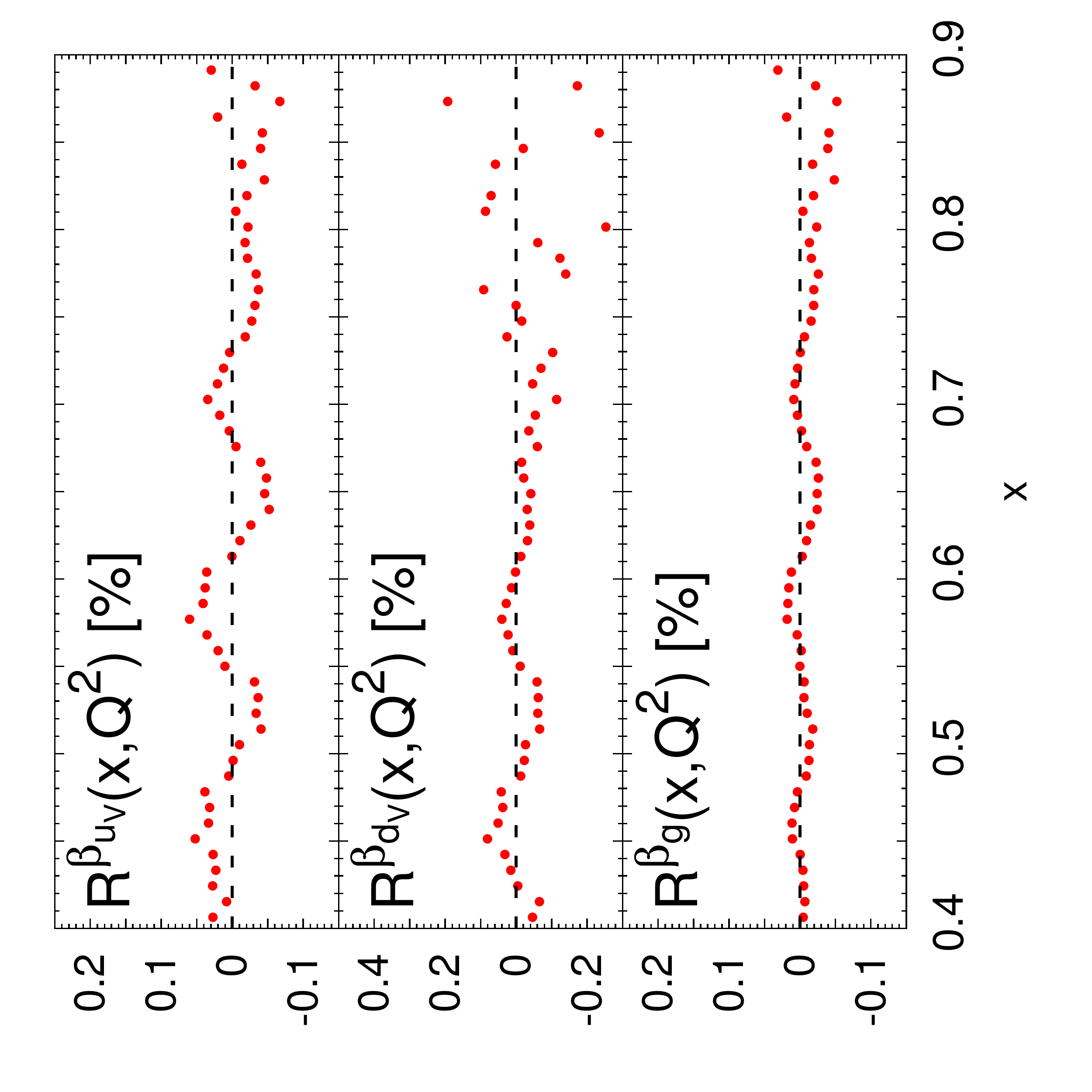}
\vspace{-0.7cm}
\caption{\small The percentage difference $R(x,Q^2)$, Eq.~(\ref{eq:Rperc}), between
the numeric and analytic evaluation of the effective 
exponents $\alpha_{f_i}$ (left) and $\beta_{f_i}$ (right)
for the {\tt MSTW08} NLO set at $Q^2=1$ GeV$^2$.}
\label{fig:anavsnum}
\end{figure}

The robustness of our numerical procedure can be validated by comparing it
with an analytic evaluation of Eq.~(\ref{eq:def}). For instance, we consider 
the {\tt MSTW08} NLO PDF set at $Q^2=1$ GeV$^2$, 
and compute the central value of 
the effective exponents $\alpha_{f_i}(x,Q^2)$ and $\beta_{f_i}(x,Q^2)$
both analytically and numerically.
The relative difference between the two computations, defined as
\begin{equation}
R^{\alpha_{f_i}}(x,Q^2) = 
\frac{\alpha_{f_i}^{\rm (num)}(x,Q^2) - \alpha_{f_i}^{\rm (ana)}(x,Q^2)}
{\alpha_{f_i}^{\rm (ana)}(x,Q^2)} 
\ \ \ \ \
R^{\beta_{f_i}}(x,Q^2) = 
\frac{\beta_{f_i}^{\rm (num)}(x,Q^2) - \beta_{f_i}^{\rm (ana)}(x,Q^2)}
{\beta_{f_i}^{\rm (ana)}(x,Q^2)}
\,\mbox{,}
\label{eq:Rperc}
\end{equation}
is displayed in Fig.~\ref{fig:anavsnum} for the $u_V$, $d_V$ and $g$ PDFs.
The agreement between the analytic (ana) and numeric (num) computation is
excellent: relative differences are at the permille level, with the only 
exception of $\alpha_g(x,Q^2)$ around $x\sim 10^{-3}$, where the input gluon 
PDF has a node.